\documentclass[a4paper,11pt]{article}
\pdfoutput=1 
\usepackage{jheppub}
\usepackage{amsmath,amssymb,bbm}
\usepackage{braket,ulem,booktabs}
\usepackage{graphicx}
\usepackage{hyperref}
\usepackage{enumitem}
\usepackage{stackrel}
\usepackage{graphicx}
\usepackage{subcaption}
\usepackage{tikz}
\usepackage{mathdots}
\usepackage{array}
\usepackage{bm}
\usepackage{changes}

\makeatletter
\def\SL@eqntext#1{\rlap{\quad\SL@margintext{#1}}}
\makeatother

\DeclareMathOperator{\Tr}{Tr}

\def\bne{\begin{equation}}
\def\ene{\end{equation}}
\newcommand{\del}{\partial}

\title{Krylov exponents and power spectra in the EFT of maximal chaos 
}

\title{Krylov exponents and power spectra for maximal quantum chaos: an EFT approach}

\author[a]{Saskia Demulder,}
\author[b]{Maria Knysh,}
\author[b]{and Andrew Rolph}

\affiliation[a]{Department of Physics, Ben-Gurion University of the Negev, David Ben Gurion Boulevard 1, Beer Sheva 84105, Israel}
\affiliation[b]{Vrije Universiteit Brussel (VUB) and The International Solvay Institutes, Pleinlaan 2, B-1050 Brussels, Belgium}

\abstract{
We examine the effective field theory (EFT) of maximal chaos through the lens of Krylov complexity and the Universal Operator Growth Hypothesis. We test the relationship between two measures of quantum chaos: out-of-time-ordered correlators (OTOCs) and Krylov complexity. In the EFT, a shift symmetry of the hydrodynamic modes enforces the maximal Lyapunov exponent in OTOCs, $\lambda_L = 2\pi T$, while simultaneously constraining thermal two-point autocorrelators. We solve these constraints on the autocorrelator, and calculate the Lanczos coefficients and Krylov exponents for several examples, finding both $\lambda_K = \lambda_L$ and $\lambda_K = \lambda_L/2$. This demonstrates that, within the EFT, the shift symmetry alone is insufficient to enforce maximal Krylov exponents even when the Lyapunov exponent is maximal. In particular, this result suggests a tension with the conjectured bound $\lambda_L \leq \lambda_K \leq 2\pi T$. Finally, we identify autocorrelator solutions whose power spectra closely resemble the so-called thermal product formula seen in holographic systems.
}

\begin{document}

\maketitle
\section{Introduction and motivation}

Understanding how information spreads and equilibrates in quantum many-body systems is a fundamental challenge in modern theoretical physics. 
Recent advances have focused on the time evolution and growth of operators to understand thermalisation and scrambling driven by the chaotic nature of theories. Over the years, a toolkit has developed for probing and understanding chaos in quantum many-body systems. One of these probes is the out-of-time-ordered correlator (OTOC) of two generic, normalised, few-body operators $V$ and $W$. This specific correlation function measures how early perturbations of $W$ affect later measurements of $V$, mirroring the classical butterfly effect. 

In many-body chaotic systems at finite temperature $T = \beta^{-1}$, the initial perturbation created by $W$ scrambles and the OTOC exhibits an exponential decay after the relaxation time $t_r$\footnote{In continuum field theories, one needs to regularise this expression, because of local operator coincident point singularities in the correlators. Typically, this is done by spacing out the operators around the thermal circle. 
}
\begin{align}\label{eq:otoc}
\frac{\langle V(t)W(0)V(t)W(0)\rangle_\beta}{\langle V(t)V(t)\rangle_\beta \langle W(0) W(0) \rangle_\beta} \sim 1 - \frac{a}{N}e^{\lambda_L t}\,,\qquad t_r\ll t\ll t_s\,.
\end{align}
with $N \gg 1$ a measure of the number of degrees of freedom, and $a$ a theory-dependent order $1$ constant. The OTOC decays until the scrambling time $t_s \sim \lambda_L^{-1} \log N$, and the decay rate is controlled by the Lyapunov exponent $\lambda_L$. The Lyapunov exponent is upper-bounded by the Maldacena-Shenker-Stanford (MSS) bound \cite{Maldacena:2015waa}%
\footnote{We work in natural units, $k_B = \hbar = 1.$}%
\begin{align}\label{eq: MSS bound}
    \lambda_L\leq 2\pi T\,.
\end{align} 
Systems that saturate this bound are referred to as maximally chaotic. Examples include holographic models, such as the Sachdev-Ye-Kitaev (SYK) model in the large $N$, low temperature limit~\cite{Shenker:2013pqa, Roberts:2014ifa, Maldacena:2016hyu}.

To capture this universal feature of chaotic dynamics in these systems, an effective field theory (EFT) approach has been developed in \cite{Blake:2017ris, Blake:2021wqj}. The key assumption is that the growth of generic few-body operators comes from their coupling to hydrodynamic modes. The construction relies on building an action for these hydrodynamic modes to all orders in a derivative expansion, referred to as quantum hydrodynamics, which is based on symmetry principles following the methods outlined in \cite{Crossley:2015evo, Glorioso:2017fpd}. In addition to the symmetries needed for quantum hydrodynamics, a crucial structural ingredient in this EFT construction is a postulated shift symmetry, which acts on the hydrodynamic variables and constrains the allowed interactions. This shift symmetry was shown to ensure the growth of the OTOCs in $0+1$ dimensional systems \cite{Blake:2017ris}. Note that while this EFT successfully captures key features of maximally chaotic dynamics as probed by the OTOC, it does not, for instance, describe the full microscopic structure of operator growth, nor does it generalise straightforwardly to dynamics beyond leading order in $1/N$. 

Complementing OTOCs, the Universal Operator Growth Hypothesis (UOGH) and the associated Krylov complexity have emerged as diagnostics of operator growth and quantum chaos. See \cite{Parker:2018yvk} for the original proposal, \cite{Nandy:2025ktk,Baiguera:2025dkc,Rabinovici:2025otw} for comprehensive reviews, and \cite{Caputa:2021sib,Rabinovici:2022beu,Erdmenger:2023wjg,Hashimoto:2023swv, Camargo:2023eev,Camargo:2024deu,Alishahiha:2024vbf, Craps:2024suj,Malvimat:2024vhr,Chattopadhyay:2024pdj,Jian:2020qpp,Hornedal:2022pkc,He:2022ryk,Bhattacharjee:2022vlt,Baggioli:2024wbz,Chen:2024imd,Ambrosini:2024sre,Caputa:2024sux} for recent work exploring their connection to chaos. Krylov complexity quantifies how an operator spreads in Krylov space, the space spanned by nested commutators of the Hamiltonian with an initial operator, constructed via the Lanczos algorithm. The resulting Lanczos coefficients $b_n$ define a tridiagonal representation of the Liouvillian, and in chaotic systems in the thermodynamic limit are conjectured to grow linearly, $b_n \sim \alpha n$ as $n \to \infty$. This linear growth yields, under certain conditions, exponential growth of the Krylov complexity, $K(t) \propto \exp(\lambda_K t)$, governed by the Krylov exponent $\lambda_K$. The latter is conjectured to upper-bound the Lyapunov exponent derived from the OTOC.

More specifically, at finite temperature, the Krylov exponent has been conjectured, and proven at infinite temperature, to provide a tighter bound on chaos than the MSS bound\footnote{Note that at finite temperature, this statement depends on the chosen regularization scheme used to define the thermal OTOC. Therefore, the Lyapunov exponent $\lambda_L$  can in general depend on this choice, see e.g.\ \cite{Romero-Bermudez:2019vej}. We will comment on this when reviewing the conjecture in detail around eq. \eqref{eq:bound_fromalpha}.}
\begin{align} \label{eq:Kchabintro}
\lambda_L \leq \lambda_K \leq 2\pi T\,.
\end{align}
Supporting evidence for this conjecture was given in \cite{Avdoshkin:2019trj, Gu:2021xaj}. However, the bound also highlights a limitation of Krylov complexity as a diagnostic for chaos. In particular, systems with maximal $\lambda_K$ and submaximal $\lambda_L$ have been constructed \cite{Chapman:2024pdw, Bhattacharjee:2022vlt}, indicating that Krylov complexity can, in some cases, fail to detect the submaximal and even integrable nature of the model.

More issues with this bound come from quantum field theory (QFT) considerations. Specifically, correlation functions of generic operators in QFTs exhibit short-distance singularities when the operator insertions approach the same spacetime point, known as the coincident point limit. For a smooth Lanczos sequence, such singularities in the two-point function fully determine the asymptotes of the Lanczos coefficients. For example, in \cite{Dymarsky:2021bjq}, maximal linear growth of the Lanczos sequence was obtained for integrable and free field theories, indicating that this behaviour should not always be associated with the presence of chaos. However, by introducing an energy cutoff, e.g. a UV cutoff, the value of $\lambda_K$ can distinguish integrable and chaotic field theories
\cite{Camargo:2022rnt, Avdoshkin:2022xuw}.

An important observation towards the central aim of this paper is that when restricting to maximally chaotic systems with $\lambda_L=2\pi T$, the bound in eq.~\eqref{eq:Kchabintro} reduces to stating that $\lambda_K=2\pi T$. The EFT provides an interesting framework to test this statement since by construction, the description only captures IR degrees of freedom and thus naturally features a UV cutoff. We will consider $0+1$ dimensional systems since the EFT is sharply formulated in this case and the OTOCs are shown to always feature maximal exponential growth $\lambda_L = 2\pi T$.

Our strategy for examining whether maximal chaos as seen through the OTOC implies maximal Krylov growth in the EFT is to find two-point functions that satisfy the shift symmetry constraints, as well as a minimal set of physical requirements: KMS symmetry, reality, and unitarity. For these solutions, we then compute the corresponding Lanczos coefficients. In the case where these coefficients are a smooth function of $n$, when extended from the integers to the reals, the asymptotic growth of the found $b_n$ determines the associated Krylov exponent $\lambda_K$~\cite{Parker:2018yvk}.
A simple example that solves the shift symmetry constraints is the conformal correlator $g_{\hat{V}}(t) = \sinh^{-2\Delta_V} (\pi t/\beta)$. Its Lanczos coefficients are known: they are asymptotically linear, $b_n \sim \frac{\pi}{\beta}n$, giving $\lambda_K = 2\pi T = \lambda_L$ and thus this example satisfies the UOGH and the exponent bound eq.~\eqref{eq:Kchabintro}. However, not all solutions of the shift symmetry constraints that we construct satisfy the exponent bound. In particular, we find that, in the considered examples, the Krylov exponent $\lambda_K$ takes one of two values: $2\pi T$ or $\pi T$. The latter case, $\lambda_K=\pi T$, suggests that the shift symmetry is not enough to ensure maximal Krylov growth within our working assumptions. Moreover, we will comment on the possible tension with eq.~\eqref{eq:Kchabintro}, as well as with the argument for the bound given in~\cite{Gu:2021xaj} for a restricted class of models. However, as we will discuss in~\autoref{sec:interpr}, the argument of \cite{Gu:2021xaj} relies on a specific factorisation structure of the OTOC, which is not assumed in the EFT construction. Our result raises the possibility that the conjecture may be more sensitive to dynamical details than previously appreciated. 

This paper is organised as follows. In \autoref{sec:background}, we start by reviewing the effective field theory (EFT) description of chaos proposed in \cite{Blake:2017ris, Blake:2021wqj}. We also include a summary of the UOGH and the definition of the Krylov exponent. Building on this, we summarise the various constraints that are to be imposed on two-point functions of generic operators that arise in this framework. This includes the main ingredient of the EFT construction, a constraint on two-point functions due to the presence of a shift symmetry. In \autoref{sec:Krylov_vs_EFT}, we solve the shift symmetry constraints for several illustrative examples. Computing the corresponding Krylov exponent, we encounter cases that exhibit both maximal and submaximal growth. Finally, we provide a critical assessment of these results, contrasting the defining assumptions of the EFT and Krylov complexity. In \autoref{sec:solv_shift}, we turn to a general analysis of the solutions of the shift symmetry constraints and derive a thermal product-like formula for the power spectrum for a set of examples. We show that the solution space of the shift symmetry constraints can be identified with a subclass of Meijer G-functions. Finally, we complement the analysis in the main text with several technical appendices.

\section{Operator growth in the EFT for many-body maximal quantum  chaos}\label{sec:background}
In this section, we review both the effective field theory approach to maximally chaotic many-body quantum systems, as proposed by \cite{Blake:2017ris,Blake:2021wqj}, and the UOGH and Krylov complexity. We also explain the constraints that are to be imposed on two-point functions of generic operators in the EFT.

\subsection{The hydrodynamical EFT of maximal chaos}
\label{sec:backgroundEFT}
In \cite{Blake:2017ris, Blake:2021wqj}, the authors developed a hydrodynamic effective theory for maximally chaotic quantum systems, capturing the universal features of operator scrambling in large-$N$ many-body systems. Let us start by briefly reviewing the basic ingredients of this formalism. 

\subsubsection*{Effective field theory on a CTP contour}
To describe the real-time dynamics of quantum systems, especially out of equilibrium, one must adopt a closed-time path (CTP) approach. Schwinger-Keldysh effective field theory provides a framework to do just so. It computes expectation values by evolving forward and then backwards along a time contour. This naturally leads to an integral path defined over two time branches, doubling the dynamical fields and their sources.  

More precisely, the Schwinger-Keldysh generating functional of connected correlators is defined as
\begin{equation}
e^{W[A_1,A_2]}= \int_{\rho} {\mathcal D}\psi_1{\mathcal D}\psi_2\, e^{iI(\psi_1;\,A_1)-iI(\psi_2;\,A_2)},
\label{eqn: SK micro}
\end{equation}
where $I$ is the functional capturing the dynamics of the microscopic degrees of freedom $\psi_1, \psi_2$  along the forward and backwards segments of the contour, respectively, with corresponding external sources $A_1, A_2$, and $ \rho$ is the initial density matrix.

Upon integrating out high-energy modes, one obtains an effective theory for low-energy degrees of freedom $\Phi_1, \Phi_2$, with
\begin{equation}
e^{W[A_1,A_2]}= \int {\mathcal D}\Phi_1{\mathcal D}\Phi_2\, e^{iI_\mathrm{eff}(\Phi_1,\Phi_2;A_1,A_2;\rho)}.
\label{eqn: SK EFT}
\end{equation}
Here, $I_\mathrm{eff}$ is a functional of the doubled fields and sources, as well as of the initial state. 

The most general action that can come from such a procedure has to be consistent with the symmetries and physical constraints of the system. Some of these constraints, such as those associated with conserved quantities, are model-dependent. Others are universal and follow from fundamental principles like unitarity and CPT invariance of the underlying quantum theory.  Moreover, if the initial state is thermal, $\rho_\beta = Z(\beta)^{-1} e^{-\beta { H}}$, where ${ H}$ is the Hamiltonian, the effective action also satisfies the Kubo-Martin-Schwinger (KMS) symmetry. See \cite{Crossley:2015evo, Glorioso:2017fpd} for more details on this construction and \cite{Liu:2018kfw} for a review. The explicit expression for the most general action consistent with these requirements is rather intricate and does not play a direct role in the arguments presented in this paper. We therefore refer the reader to \cite{Blake:2017ris} for further details. 

\subsubsection*{Shift symmetry of the hydrodynamic action in $0+1$D}

In quantum many-body systems where energy is the only conserved quantity, long-time dynamics are governed by a single hydrodynamic mode. This mode, denoted by $\epsilon(t)$, describes local fluctuations in time evolution, and appears as a small reparametrisation of the physical time coordinate: $t \mapsto t + \epsilon(t)$.
Physically, this mode captures fluctuations around thermal equilibrium in a way consistent with energy conservation.

The dynamics of $\epsilon(t)$ are governed by an effective action, constructed in the Schwinger-Keldysh formalism as sketched above. In systems with a large number $N$ of degrees of freedom, this action scales as $O(N)$, while correlation functions of the hydro mode are suppressed by powers of $1/N$. For example, the two-point function is of the order of $1/N$.

A crucial input in \cite{Blake:2017ris, Blake:2021wqj} to render the effective action maximally chaotic is requiring that the effective action of the hydrodynamic mode is invariant under a shift symmetry
\begin{equation} \label{eq:shiftsymm}
\epsilon(t) \mapsto \epsilon(t) + a_+ e^{\lambda_L t} + a_- e^{-\lambda_L t},
\end{equation}
for arbitrary constants $a_\pm$, with $\lambda_L = 2\pi/\beta$, where $\beta$ is the equilibrium inverse temperature, the maximal Lyapunov exponent. 

\subsubsection*{Exponentially growing OTOCs and Lyapunov exponents}
\begin{figure}[t]
    \centering
    \includegraphics[width=0.7\linewidth]{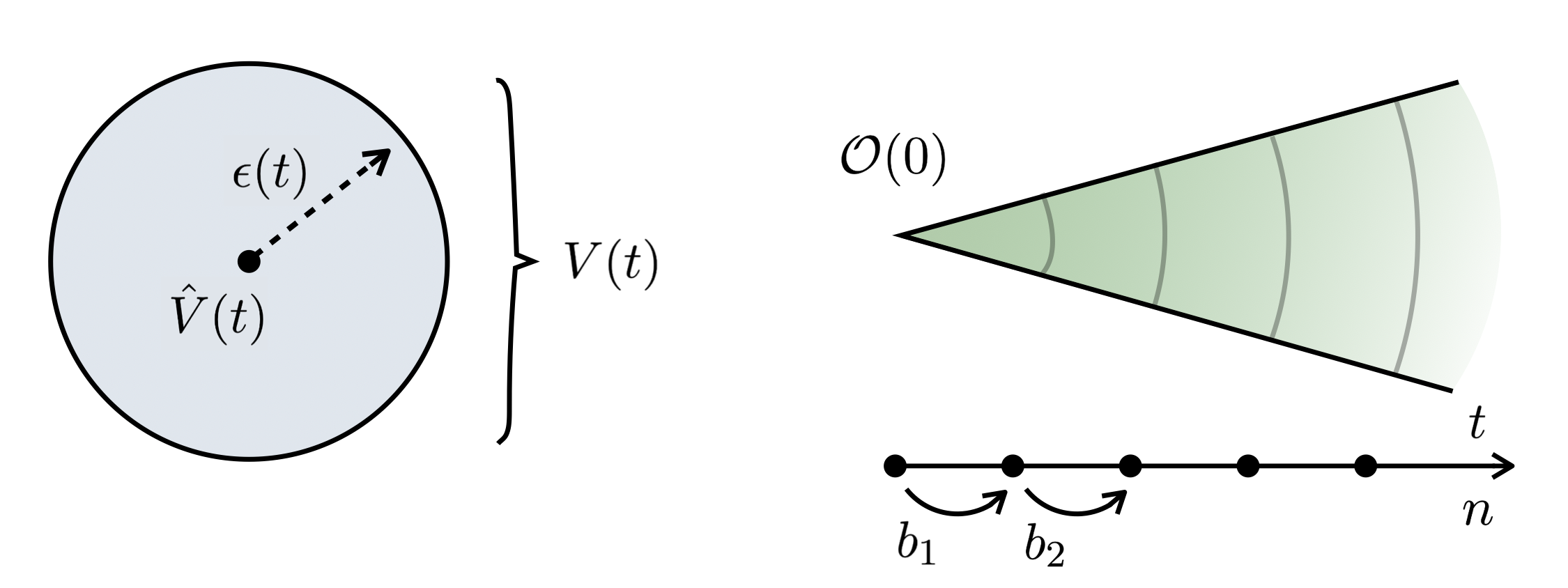}
    \caption{Two perspectives on operator growth. Left: spread of the operator cloud in the space of degrees of freedom, driven by the hydro mode $\epsilon$. Right: spreading across the Krylov basis in Hilbert space, governed by the Lanczos sequence $\{b_n\}$. We will identify the initial operator $\mathcal O(0)$ with the bare field operator in the EFT.}
    \label{fig:EFT_vs_Krylov}
\end{figure}

 The framework of \cite{Blake:2017ris, Blake:2021wqj} further builds on the intuition that the evolution of a generic few-body bosonic operator $V(t)$ at finite temperature can be decomposed into a ``bare'' component $\hat{V}(t)$, which captures its microscopic degrees of freedom, and a hydrodynamic ``cloud'' arising from coupling to a hydro mode $\epsilon$ associated with energy conservation
\begin{align}
\label{eq: Vexpansion}
	V(t) = \hat{V}(t) + L^{(1)}_t[\hat{V}(t), \epsilon (t)] + O(\epsilon^2)\,,
\end{align}
where $L_t^{(1)}$ is a differential operator
\begin{align}
\label{eq: Vcoupling}
	L^{(1)}_t[\hat{V}(t), \epsilon (t)] = \sum_{n,m=0}^\infty c_{nm} \partial_t^n \hat{V} (t)\, \partial_t^m \epsilon(t) \,,
\end{align}
with coefficients $c_{nm}$ encoding the effective low-energy dynamics. This decomposition is depicted in the left panel of \autoref{fig:EFT_vs_Krylov}. Crucially, \cite{Blake:2017ris, Blake:2021wqj} assume that bare operators $\hat{V}$, $\hat{W}$ that involve different degrees of freedom can only communicate with themselves: $\langle \hat{V}(t_1) \hat{W}(t_2) \rangle_\beta = 0$ for $V \neq W$.

Within this setup, two-point functions of $V(t)$ can be computed perturbatively in $\epsilon$ and using that $n$-point functions of $\epsilon$ are suppressed by powers of $1/N$, this is equivalent to expanding in $1/N$. At leading order in $1/N$, the correlation functions of $V$ and $\hat{V}$ are the same
\begin{align}\label{eq:full_to_bare_g}
    \langle V(t_1) V(t_2) \rangle_\beta = \langle \hat{V} (t_1) \hat{V} (t_2)\rangle_\beta + O(1/N)\,.
\end{align}
The thermal Wightman two-point functions of the bare operators are
\begin{align}
	g^-_{\hat{V}}(t) := \langle \hat{V}(0) \hat{V}(t) \rangle_\beta \,, \qquad g^+_{\hat{V}}(t) := \langle \hat{V}(t) \hat{V}(0) \rangle_\beta \,.
\end{align}
These satisfy physical constraints such as the KMS condition and unitarity. See \autoref{app:phys_props} for a list of physical properties of thermal correlators. Henceforth, we will work exclusively with the first thermal Wightman two-point function, and denote it simply by $g_{\hat V }\equiv g^-_{\hat V}$.

Turning to four-point functions, the key result of \cite{Blake:2017ris, Blake:2021wqj} is that at leading order $\epsilon (t)$, the OTOCs can be expressed in terms of effective vertices coupling two bare operators to the hydrodynamic $B_{\hat{V}}$
\begin{align}
\label{eq:OTOCeffB}
\frac{\langle V(t_1)V(t_2)W(t_3) W(t_4)\rangle_\beta}{\langle V(t_1)V(t_2)\rangle_\beta \langle W(t_3) W(t_4) \rangle_\beta} - 1 = \langle B_{\hat{V}}(t_1,t_2) B_{\hat{W}}(t_3,t_4)\rangle,
\end{align}
where
 \begin{align}
 \label{eq: Vvertex}
 	B_{\hat{V}}(t_1,t_2) = L^{(1)}_{t_1}[g_{\hat{V}}(t_{12}) \epsilon(t_1)] + L^{(1)}_{t_2}[g_{\hat{V}}(t_{12}) \epsilon(t_2)] \,.
 \end{align}

A central feature of the construction is an additional constraint on the form of the effective coupling, motivated by the shift symmetry. Specifically, requiring that the vertex coupling remains invariant under the exponential shifts in eq.~\eqref{eq:shiftsymm} leads to the following constraint on thermal two-point functions:
\begin{align}
 \label{eq:shifteq}
 	L^{(1)}_{t_1}[g_{\hat{V}}(t_{12})\, e^{\pm\lambda_L t_1}] + L^{(1)}_{t_2}[g_{\hat{V}}(t_{12})\, e^{\pm\lambda_L t_2}] = 0\,.
 \end{align}
These constraints will be frequently used in what follows, and we will refer to them as the ``shift symmetry constraints'' on two-point functions.

As shown in \cite{Blake:2017ris, Blake:2021wqj}, the shift symmetry together with the constraints in eq.~\eqref{eq:shifteq} are the essential ingredients in the effective theory construction that ensure exponential growth appears only in OTOCs, and not in time-ordered correlators (TOCs). In TOCs, the symmetry enforces a cancellation of the growing contributions, so they do not exhibit any exponential sensitivity to time separations. In contrast, the non-trivial time ordering in OTOCs obstructs this cancellation, leading to exponential growth in the connected piece of the OTOC, as in eq.~\eqref{eq:otoc}. In this way, the hydrodynamic effective theory of \cite{Blake:2017ris,Blake:2021wqj} reproduces the universal hallmark of maximal quantum chaos: OTOCs grow exponentially in time with $\lambda_L=2\pi T$, while TOCs do not grow exponentially. In this paper, unless otherwise stated, $\lambda_L$ always denotes its value in the EFT: $\lambda_L = 2\pi T.$

\subsection{The Universal Operator Growth Hypothesis and Krylov exponents }\label{eq:KC_UOGH}

Krylov complexity has recently emerged as another powerful probe of quantum chaos. It quantifies how rapidly an operator spreads in the Hilbert space under Heisenberg time evolution, as measured in the Krylov basis generated by repeated commutators with the Hamiltonian. In chaotic systems, this spread is expected to be exponential, paralleling the behaviour of OTOCs.

The Krylov complexity of a time-evolving operator $\hat{V}(t)$ can be calculated  from its thermal autocorrelator
\begin{align}\label{eq:Ct_wightman}
    C(t) = \operatorname{Tr}\left[\rho_\beta^{1/2} \hat{V}(t) \rho_\beta^{1/2} \hat{V}(0)\right] \,,
\end{align}
with $\rho_\beta = e^{-\beta H}/Z$ the thermal density matrix and $V(t)=e^{iHt} V(0) e^{-iHt}$. This correlator is related to the thermal Wightman function via
\begin{align}
    C(t) = g_{\hat{V}}\left(t + \tfrac{i\beta}{2}\right) \,, \qquad \text{Im } t \in \left[-\tfrac{\beta}{2}, \tfrac{\beta}{2} \right] \,.
\end{align}

A proposal to quantify operator growth was put forward in \cite{Parker:2018yvk}, where the idea is to study the evolution of $\hat{V}(t)$ by constructing a Krylov basis from the nested commutators $\{\mathrm{ad}_H^n \hat{V}\}_n$.  Using the Lanczos algorithm, a variation of the Gram-Schmidt orthogonalisation specialised to Hermitian dynamics,  one obtains an orthonormal basis $\{|\hat{V}_n)\}_n$ and a sequence of real positive Lanczos coefficients $\{b_n\}$. In this basis, the dynamics of the operator are equivalent to a tight-binding chain with variable hopping
\begin{align}\label{eq:tightb}
    \partial_t \varphi_n(t) = -b_{n+1} \varphi_{n+1}(t) + b_n \varphi_{n-1}(t) \,,
\end{align}
where $\varphi_n(t) = i^{-n} (\hat{V}_n | \hat{V}(t))$ encodes the amplitude of the operator in the $n$-th Krylov basis vector. The operator's growth across the basis is thus governed entirely by the Lanczos sequence. Commonly used methods to compute the Lanczos coefficients are reviewed in \autoref{app:toda}.

\subsubsection*{Lanczos coefficients and the UOGH}
The central conjecture of \cite{Parker:2018yvk}, the UOGH, is that, in quantum chaotic many-body systems, the Lanczos coefficients grow as fast as possible. Specifically, the coefficients grow linearly in $n$ at asymptotically large values of $n$
\begin{align}\label{eq:bn_asympt}
    b_n \sim \alpha n   + \gamma + o(1), \qquad n\to \infty\,,
\end{align}
leading to exponential growth of the Krylov complexity
\begin{align}
    K_{\hat{V}}(t) = \sum_n n |\varphi_n(t)|^2
    \underset{t \to \infty}{\sim} e^{\lambda_K t}\,.
\end{align}

Here, $\varphi_n$ are the wave-functions obtained by solving the auxiliary one-dimensional tight-binding chain in eq.~\eqref{eq:tightb}.
The rate $\lambda_K$ is known as the Krylov exponent and characterises the effective speed of operator spreading. 
Note that the value of the Lanczos coefficients depends both on the choice of inner product and on the temperature. In what follows, the inner product is always taken to be the Wightman inner product defined in eq.~\eqref{eq:Ct_wightman}, while the temperature dependence of the coefficients will be left implicit.

The authors in \cite{Parker:2018yvk} conjectured that the Krylov exponent provides a bound on chaos, proposing that the maximal rate of operator growth bounds the maximal rate of scrambling as measured by the OTOC\footnote{
The original conjecture as stated in \cite{Parker:2018yvk} considers the thermal OTOCs as defined in \cite{Maldacena:2015waa} which are of the form $\Tr\!\left[y\,V(0)\,y\,W(t)\,y\,V(0)\,y\,W(t)\right]$ with $y^4=\rho_\beta$. In general, the Lyapunov spectrum extracted from the thermal OTOCs can depend on the choice of regularization scheme. An example where such contour dependence was considered in weakly coupled field theories is \cite{Romero-Bermudez:2019vej}. On the other hand, in large-$N$ models such as SYK, the Lyapunov spectrum does not exhibit contour dependence \cite{Shenker:2014cwa,Chowdhury:2017jzb,Romero-Bermudez:2019vej}, consistent with the expectation that in maximally chaotic systems the exponent is universal. In this work, we will extract $\lambda_L$ from the thermal OTOC used in \cite{Blake:2017ris}, which of the form $\text{Tr}\!\left[\rho_\beta\,V(0)\,W(t)\,V(0)\,W(t)\right]$, and interpret comparisons to $\lambda_K$ accordingly. \label{footnote4}}
\begin{align}\label{eq:bound_fromalpha}
    \lambda_L\leq 2\alpha\,.
\end{align}
If the Lanczos coefficients asymptote to $b_n \sim \alpha n$ as $n \to \infty$, and when the associated Krylov complexity is exponentially growting, we can identify $2\alpha=\lambda_K$. If not, the Krylov exponent is obtained by solving~\eqref{eq:tightb} and computing the Krylov complexity directly. At infinite temperature, this conjectured relation between $\alpha$ and the Lyapunov exponent has been proven~\cite{Parker:2018yvk}. At finite temperature, the Lanczos coefficients, and hence their slope $\alpha$, acquire a temperature dependence. In \cite{Parker:2018yvk}, by adopting the Wightman inner product, the authors conjecture that the bound in eq.~\eqref{eq:bound_fromalpha} still applies and is verified in several examples, including Schwarzian quantum mechanics and the large-$N$, large-$q$ limit of SYK (taking $N$ large first).  In particular, in the former, the Krylov exponent was shown to saturate the bound: $\lambda_L = \lambda_K$.

\subsubsection*{A power spectrum version of the UOGH}
Another insightful perspective on the Krylov exponent is that it captures the decay properties of the power spectrum at large frequencies. Indeed, the large-frequency behaviour of the power spectrum directly encodes the analytic structure of the autocorrelator in the complex plane. This relationship is explicitly realised through the Fourier transform
\begin{align}\label{eq:Phi_to_Ct}
\Phi(\omega) = \int_{-\infty}^{+\infty} C(t) e^{i\omega t}\,\mathrm{d}t\,,
\end{align}
assuming that $C(t)$ has a well-defined Fourier transform.
We also define the moments of the autocorrelator or power spectrum as follows
\begin{align}\label{eq:moments}
    \mu_{2n}  = (-1)^n\left. \frac{d^{2n}}{dt^{2n}} \, C(t) \right|_{t=0}\,,\qquad \mu_{2n} = \frac{1}{2\pi} \int_{-\infty}^\infty \omega^{2n} \Phi(\omega) d\omega \,.
\end{align}

Assuming the power spectrum shows an exponential decay of the form
\begin{align}\label{eq:decay_Phi_gen}
    \Phi(\omega)\sim e^{-(\omega/\omega_0)^{2/\rho}}\,, \qquad \omega \to \infty\,,
\end{align}
the Lanczos coefficients display an asymptotic behaviour of the form $b_n^2\sim n^\rho$ \cite{lubinsky1993update}, where the proportionality constant is related to the decay rate $\omega_0$ in eq.~\eqref{eq:decay_Phi_gen}. For linear growth  ($\rho = 2$), the asymptotic growth rate of the Lanczos coefficients $b_n$ in eq.~\eqref{eq:bn_asympt} and exponential decay rate of the power spectrum $\Phi(\omega)$ are related as  $\omega_0 = 2\alpha/\pi$. 

From the above, we have retrieved, provided the smoothness condition on the Lanczos coefficients is satisfied, the known cycle of relations between the autocorrelator, power spectrum, and Lanczos coefficients
\begin{align}
\raisebox{-30pt}{\begin{tikzpicture}[x=0.75pt,y=0.75pt,yscale=-1,xscale=1]
\draw (177,54.4) node [anchor=north west][inner sep=0.75pt]    {$C( t)$};
\draw (214,94.4) node [anchor=north west][inner sep=0.75pt]    {$\Phi ( \omega )$};
\draw (150,94) node [anchor=north west][inner sep=0.75pt]    {$
b_{n}$};
\draw (225,70) node [anchor=north west][inner sep=0.75pt]    {\eqref{eq:Phi_to_Ct}};
\draw (174,110) node [anchor=north west][inner sep=0.75pt]    {\eqref{eq:moments}};
\draw (120,70) node [anchor=north west][inner sep=0.75pt]    {\eqref{eq:dym}};
\draw (177,100.4) node [anchor=north west][inner sep=0.75pt]    {$\longleftrightarrow $};
\draw (155,90) node [anchor=north west][inner sep=0.75pt]  [rotate=-310]  {$\longleftrightarrow $};
\draw (220.76,95) node [anchor=north west][inner sep=0.75pt]  [rotate=-235]  {$\longleftrightarrow $};
\draw (274,99.4) node [anchor=north west][inner sep=0.75pt]    {.};
\end{tikzpicture}}\label{tikz:relsKrylov}
\end{align}
In particular, the analyticity properties of $C(t)$ are related to the properties of the other corners of the triangle, most notably; linear growth of $b_n$ implies that $C(t)$ is analytic in a strip of width $1/\omega_0$ in the complex time plane, with $\omega_0 = 2\alpha/\pi$ \cite{Parker:2018yvk}. It is important to emphasise that the relation $\pi \omega_0 = 2\alpha = \lambda_K$ only holds when $b_n$ is asymptotically linear and when the associated Krylov complexity grows exponentially. When the Lanczos coefficients $b_n$ are not smooth functions of $n$, when $n$ is extended from the integers to the reals, the simple correspondence may no longer hold, and this was explored in \cite{Dymarsky:2021bjq,Camargo:2022rnt,Avdoshkin:2022xuw,Anegawa:2024wov,Anegawa:2024yia}. Persistent staggering, defined in~\cite{Avdoshkin:2022xuw} as when $b_n$ is asymptotically linear but with different intercepts for odd and even $n$, is an example of non-smooth behaviour that has been studied.

\subsubsection*{A conjectured tighter bound on the Lyapunov exponent}

At finite temperature, using the relation between the slope of the Lanczos coefficients and the high-frequency decay of the spectral function, \cite{Parker:2018yvk} showed that the decay rate $\alpha$ satisfies the inequality
\begin{align} \label{eq:alpha_ub}
\alpha \leq \pi T\,,
\end{align}
provided that the thermal Wightman spectral function decays at least as fast as $e^{-\beta \omega/2}$ at large $\omega$. Combining this with the bound $\lambda_L \leq 2\alpha$ yields a potentially stronger constraint on the Lyapunov exponent than the MSS bound.

The applicability of this bound relies on the analytic properties of the two-point function and, crucially, on the behaviour of the Lanczos coefficients. In particular, it assumes that the coefficients $b_n$ are asymptotically linear in $n$ with slope $\alpha$, in which case $2\alpha = \lambda_K = \pi \omega_0$~\cite{Parker:2018yvk}. However, when the Lanczos coefficients are not smooth, such as when they display staggering, this correspondence breaks down. In such cases, $\lambda_K$ can still be defined directly from the growth of Krylov complexity $K(t)$. This motivates a refined version of the quantum chaos bound that generalises eq.~\eqref{eq:alpha_ub} and remains applicable even when $b_n$ is non-smooth:
\begin{align} \label{eq:conjecture}
\lambda_L \leq \lambda_K \leq 2\pi T \,.
\end{align}
Let us stress that this inequality should be interpreted with care, as highlighted in \autoref{footnote4}.
Further support for this type of bound, under certain technical assumptions, was later provided by \cite{Gu:2021xaj}, who proposed $\lambda_L \leq \pi \omega_0 \leq 2\pi T$, see also \cite{Avdoshkin:2019trj}. 

\subsection{More on the shift symmetry and physical conditions} \label{sec:shift_time_freq_C_phi}
So far in this section, we have given a broad review of the EFT and of Krylov complexity. In what follows, we will reformulate the EFT shift symmetry constraint in terms of the autocorrelator and the power spectrum.   To do so, we assume the hydrodynamic coupling of operators to the chaos mode as encoded in eq.~\eqref{eq: Vexpansion}, and take the autocorrelator of the bare field $\hat V$ to provide a reliable approximation to that of the full operator $V$, see eq.~\eqref{eq:full_to_bare_g}.  With those ingredients and relations at hand, we will derive the Lanczos coefficients capturing the operator growth of the bare-field operator in the EFT of maximal chaos in the next section.

\subsubsection{Shift symmetry for the autocorrelator and the power spectrum}
Recall that for the EFT to describe maximally chaotic quantum systems, the Wightman two-point function of the bare field $\hat{V}$ must satisfy the shift constraints~\eqref{eq:shifteq}, which can be equivalently written as
\begin{align}\label{eq:shift} 
\sum_{n=0}^\infty f_{n}(\pm\lambda) (e^{\pm\lambda t}+(-1)^n) \del_t^n g_{\hat{V}} (t) = 0 \,,
\end{align} 
where we use $\lambda_L=\lambda$ for simplicity of notation and
\begin{align}
    \label{eq:def_polys}
    f_{n}(\lambda)=\sum_{m=0}^\infty c_{nm}\lambda^m.
\end{align} 
The constraints can also be rewritten in terms of $C(t)$, recalling that $C(t) = g_{\hat{V}} (t+i\beta/2)$, as follows
\begin{align}
    \label{eq:shift2} \sum_{n=0}^\infty f_{n}(\pm\lambda) (e^{\pm\lambda t}+(-1)^{n+1}) \del_t^n C (t) = 0\,.
\end{align}
In the frequency domain, these constraints can be written as
\begin{align}\label{eq:fts} \sum_{n=0}^\infty f_{n}(\pm\lambda) ((-1)^{n+1}(i\omega \pm\lambda)^n \Phi (\omega\mp i \lambda) + (i \omega)^n \Phi (\omega))=0 \,, \end{align} 
where we used eq.~\eqref{eq:Phi_to_Ct} relating the power spectrum to the autocorrelator. 

To analyse the shift symmetry constraints in the frequency domain given in eq.~\eqref{eq:fts}, first, we focus on the constraint coming from requiring invariance under $\epsilon(t)\to\epsilon(t)+e^{\lambda t}$. Setting $\omega = -i\lambda z$, the constraint with the upper sign can be written as a first-order non-homogeneous linear recurrence relation
\begin{align}\label{eq:Fdi}  
\Phi(z+1) = \frac{p(z)}{p(-z-1)} \Phi(z)\,,\qquad p(z) := \sum_{n=0}^m f_n(\lambda) (\lambda z)^n\,.
\end{align} 
Note that we assumed at this point that the polynomials are of finite degree $m$ for simplicity.  A similar equation can be obtained for the lower sign. However, it is automatically satisfied if $\Phi(\omega)$ obeys~\eqref{eq:Fdi} and is invariant under $\lambda\to-\lambda$.

In terms of these polynomial functions $p(z)$, the constraint equations in the time domain can be equivalently written as
\begin{align}\label{eq:shifpintro}
    p(\lambda^{-1}\partial_t-1)(e^{\lambda t}g_{\hat V}(t))+p(-\lambda^{-1}\partial_t)g_{\hat V}(t)=0\,.
\end{align}
A similar equation can be obtained for the other sign.

In what follows, it will be convenient to express the solutions in terms of the roots of the polynomial $p(z)$. To this end, we factorise
\begin{align}\label{eq:factor_polys}
p(z) = \sum_{n=0}^m f_n( \lambda)(\lambda z)^n = \lambda^m f_m (\lambda) \prod_{i=1}^{m} (z - \alpha_i)\,,
\end{align}
where we have further assumed that the polynomials are of non-zero degree $m$ for convenience. Note that when $f_n(\lambda)$ are real, the set of roots $\{\alpha_j\}_{j=1}^m$ is closed under complex conjugation.

\subsubsection{KMS, unitarity and reality conditions}

The Wightman two-point function of the bare field $\hat{V}$ must not only satisfy the shift symmetry constraint but, in general, obey certain physical requirements. Below, we list the key physical properties that we will impose on the solution obtained from solving the shift symmetry constraints.
Further details on the physical constraints and their motivation are provided in \autoref{app:phys_props}.

The Wightman autocorrelator $C(t)$ must satisfy
\begin{align}
\text{KMS: } & \qquad C(t) = C(-t)\,,\label{eq:KMS_Ct}\\
\text{Unitarity:} &\qquad  C(i\tau) \geq 0\,, \qquad\quad\; \tau \in \left[-\tfrac{\beta}{2}, \tfrac{\beta}{2}\right], \\
\text{Hermitian operators:} & \qquad C(t)^* = C(t)\,, \qquad t \in \mathbb{R} .
\label{eq:unitarity_Ct}
\end{align}
Note that time reflection symmetry imposes the same constraint as KMS symmetry on the autocorrelation function.

Its Fourier transform, the power spectrum $\Phi(\omega)$, must in turn satisfy
\begin{align}
\text{KMS:} & \qquad \Phi(\omega) = \Phi(-\omega)\,, \label{eq:KMS_Phi} \\
\text{Unitarity:} &\qquad  \Phi(\omega) \geq 0\,, \qquad\qquad\;\, \omega \in \mathbb{R}\, , \\
\text{Hermitian operators:} & \qquad \Phi(\omega)^* = \Phi(\omega^*)\,, \qquad \omega \in \mathbb{C}\,.
\label{eq:unitarity_Phi} 
\end{align}
Finally, for convenience, when possible, we adopt the following convention for the normalisation of the power spectrum:
\begin{align}
\text{Normalisation:} & \qquad \Phi(0) = 1\,. \label{eq:normalisation_Phi} 
\end{align}

In what follows, we will impose this minimal set of physical constraints on the correlator solutions of the shift symmetry constraint equation; we will not, for example, assume that the theory is conformal or holographic.

\section{Simple shift-symmetric correlators and their Krylov exponents}\label{sec:Krylov_vs_EFT}
In this section, we focus on analysing four classes of correlation functions of the operator $\hat V(t)$ that solve the constraints discussed in \autoref{sec:shift_time_freq_C_phi} via their Lanczos coefficients to test the UOGH and the proposed inequality~\eqref{eq:conjecture}. 
The primary question we address here is whether the Lanczos coefficients are asymptotically linear, in line with the UOGH~\eqref{eq:bn_asympt}, and whether the shift symmetry constraints ensure that the Krylov exponent is maximal $\lambda_K = 2\pi T$.

\subsection{First order shift symmetry constraints: the Schwarzian correlator} \label{sec:Schwarzian}

A simple case one can study is for the coupling of $\hat{V}$ to the hydro mode $\epsilon$ to be such that $f_i(\lambda) = 0$ for $i>1$, and if either $f_0(\lambda)$ is an even function and $f_1(\lambda)$ odd, or $f_0(\lambda)$ odd and $f_1(\lambda)$ even. Then the shift symmetry constraints in eq.~\eqref{eq:shift} simplify to two first-order differential equations:
\begin{align}
    f_0(\lambda) (e^{\lambda t} +1) g_{\hat{V}} (t) + f_1 (\lambda)(e^{\lambda t}-1) \del_t g_{\hat{V}} (t) &=0\,, \label{eq:fiosh}\\
 f_0(\lambda) (e^{-\lambda t} +1) g_{\hat{V}} (t) - f_1 (\lambda)(e^{-\lambda t}-1) \del_t g_{\hat{V}} (t) &=0\,. \label{eq:fiosh2}
\end{align}

 An example of a theory for which the shift symmetry constraint equations are first order is Schwarzian quantum mechanics, for which the couplings between the hydro mode and the matter fields are $c_{01} = \Delta_V$ and $c_{10} = 1$~\cite{Blake:2017ris}, giving $f_0(\lambda) = \lambda \Delta_V$ and $f_1(\lambda) = 1$, see~\eqref{eq:def_polys}, and, for these $f_i(\lambda)$, the solution to eq.~\eqref{eq:fiosh2} is the Schwarzian correlator
\begin{align}\label{eq:SchwarzianCorr}
    g_{\hat{V}}(t) = \sinh^{-2\Delta} (\lambda t/2)\,.
\end{align}

Note that the map from a set of coefficients $c_{nm}$ determining the function $f_n(\lambda)$ to a particular correlator is many-to-one. For example, any pair $c_{0,j}= a \Delta_V$ and $c_{0,j+1} = a$ for any $j$, with the other $c_{nm}$'s zero, will lead to the Schwarzian correlator.

The solution to eqs.~\eqref{eq:fiosh} and~\eqref{eq:fiosh2} with general non-zero $f_0$ and $f_1$ is  
\begin{align}\label{eq:gen_sols_N=1}
	g_{\hat V}(t)=\sinh\left(\frac{\lambda t}{2}\right)^{-\frac{2 f_0(\lambda)}{\lambda  f_1(\lambda)}}\,,  
\end{align}
up to a normalisation factor.
Performing the shift $t\rightarrow t+i\beta/2$
to obtain the associated autocorrelator, and subsequently solving the recursion relation~\eqref{eq:dym} for the general solution in eq.~\eqref{eq:gen_sols_N=1} gives the following Lanczos coefficients
\begin{align}
	b_n^2 = \frac{\pi^2}{\beta^2} (n+1)\left(n+ \frac{f_0(\lambda) \beta}{f_1 (\lambda) \pi}\right)\,.
\end{align}
For large $n$, these coefficients $b_n$ are asymptotically linear in $n$: $b_n \sim \frac{\pi}{\beta}n$. Note that the asymptotic slope $\alpha$ is $\pi T$ and independent of $f_0$ and $f_1$. Also, these coefficients are a smooth function of $n$, so the asymptotic slope of $b_n$ determines the Krylov exponent, $\lambda_K = 2\pi T$. This value of $\lambda_K$ is maximal as expected for maximally chaotic systems. Hence, the solutions to eqs.~\eqref{eq:fiosh} and~\eqref{eq:fiosh2} always give a $\lambda_K$ consistent with eq.~\eqref{eq:conjecture}, for any $f_0$ and $f_1$.

\subsection{Second order shift symmetry constraints}
Going one order higher and considering the equation for $f_i(\lambda) = 0$ for $i>2$, is particularly insightful, as we will see that the shift symmetry naturally leads to two families of autocorrelators, including a correlator that is in tension with the bound (2.21). At this order, consider the case where $f_0$ and $f_2$ are even functions of $\lambda$ while $f_1$ is odd. In this case, the correlator satisfying the second-order shift symmetry constraints is
\begin{gather}\label{eq:m2_autocorr}
\begin{aligned}
    g_{\hat V} (t) &=  c_1\, e^{\frac{\eta-\kappa}{2} \lambda t}  \,
	_2F_1\left(\eta,\eta-\kappa;1-\kappa;-e^{ \lambda t }\right)  +c_2\,
	e^{\frac{\eta+\kappa}{2} \lambda t} \, _2F_1\left(\eta,\eta+\kappa;1+\kappa;-e^{ \lambda t}\right)\,,
\end{aligned}
\end{gather}
with $c_1$ and $c_2$ arbitrary constants, up to normalisation, and
\begin{align}\label{eq:antisymm_symm}
    \kappa :=\frac{\sqrt{f_1^2-4 f_0 f_2}}{ f_2 \lambda} \,,\qquad \eta :=\frac{f_1}{ f_2 \lambda} \,.
\end{align}
We analyse the possible values of $\lambda_K$ associated with this correlator. Although general analytic expressions for the Lanczos coefficients cannot be obtained from \eqref{eq:m2_autocorr} for arbitrary choices of $f_0$, $f_1$, and $ f_2$, we consider special cases for which closed-form expressions are accessible.

\paragraph{Special case 1.}
If we choose $f_1 = \lambda f_2$, and impose the KMS condition in eq.~\eqref{eq:KMS_Ct}, the autocorrelator obtained from eq.~\eqref{eq:m2_autocorr} simplifies to 
\begin{align}\label{eq:CT_spec_c1}
   C(t)  = \frac{\sinh \left(\kappa\lambda  t/2\right)}{\sinh(\lambda t/2)}\,,\qquad \kappa = \sqrt{1-\frac{4f_0}{f_2 \lambda^2}} \,,
\end{align}
up to a normalisation factor%
\footnote{Note that this correlation function $C(t)$ is not $\beta$-periodic in imaginary time since only time-ordered correlators are required to be periodic in thermal time. 
See also the massive scalar field example from~\cite{Avdoshkin:2022xuw} for an explicit $C(t)$ that is not periodic in thermal time. }. Eq.~\eqref{eq:CT_spec_c1} is analytic in the neighbourhood of $t=0$, and, for $0< \kappa < 1$, is analytic within the strip $|\text{Im} (t)| < 2\pi/\lambda = \beta$, which is twice as wide as the thermal strip.
Using the Fourier transform identity in eq.~\eqref{eq:FT_sinh/sinh} while restricting to $0<\kappa< 1$, the associated power spectrum reads
\bne
\begin{split}
\label{eq:perps}
\Phi(\omega) & =   \frac{ 1+\cos\left( \pi \kappa \right) }{ \cosh\left( \frac{2\pi \omega}{\lambda} \right) + \cos\left( \pi \kappa \right)} \,,\\
&\sim 2 (1+\cos (\pi \kappa)) e^{- \frac{2\pi \omega}{\lambda}}\,, \qquad \omega \to \infty\,.
\end{split}
\ene
We fixed the normalisation factors using the condition~\eqref{eq:normalisation_Phi}. This power spectrum is invariant under $\lambda\to-\lambda$ as required. The  Lanczos coefficients corresponding to the autocorrelator in eq.~\eqref{eq:CT_spec_c1} can be computed using the Toda-method (see~\autoref{app:toda}) and read
\begin{align}\label{eq:lanczos_f1_lambdaf2}
   b_{n-1}^2=\frac{\pi^2}{\beta^2}\frac{n^2 - \kappa^2}{4n^2-1} n^2 \,,
\end{align}
where we have set $\lambda=2\pi/\beta$, as required by consistency in the EFT~\cite{Blake:2021wqj}. In the large $n$ limit, the Lanczos coefficients behave as
\begin{align}\label{eq:lanczos_f1_lambdaf2_asympt}
    b_n \sim \frac{\pi}{2 \beta} n\,, \qquad n \to \infty \,.
\end{align}
Hence, in this case, we find 
a \textit{half-maximal} Krylov exponent, $\lambda_K = \pi T$. Note that the Lanczos coefficients in eq. \eqref{eq:lanczos_f1_lambdaf2} asymptote very quickly to a special case within the family of Lanczos coefficients associated with continuous Hahn polynomials derived in \cite{Gamayun:2025hvu}. As shown there, for these coefficients, the Krylov complexity grows exponentially, with a growth rate, in the conventions used here, equal to $\lambda_K = \pi T$. A simple numerical analysis for eq. \eqref{eq:lanczos_f1_lambdaf2} indeed confirms this.  We will defer an extensive discussion of the observation that this particular example leads to submaximal chaos to \autoref{sec:interpr}.

\paragraph{Special case 2.} 
Next, if we choose $f_1 = 0$, the roots become $\alpha_\pm =\pm \frac{i}{\lambda}\sqrt{\frac{f_0}{f_2}}$ and  the correlator in eq.~\eqref{eq:m2_autocorr} simplifies to
\bne \label{eq:f10} g_{\hat V} (t) = c_1 \sin \left(\sqrt{\frac{f_0}{f_2}} \,  t\right)+c_2 \cos
\left(\sqrt{\frac{f_0}{f_2}} \, t\right)\,. \ene

Performing the imaginary shift and imposing KMS symmetry on the resulting autocorrelator leads to 
\begin{align}\label{eq:oscillatory_sol}
   C(t)=  \cos \left(\sqrt{\frac{f_0}{f_2}}\, t \right)\,,
\end{align}
up to a normalisation constant. After Fourier transforming, the associated power spectrum and moments, obtained via eq.~\eqref{eq:moments}, are given by 
\begin{align}
\Phi(\omega) =   \delta\left(\omega - \sqrt{\frac{f_0}{f_2}}\right) + \delta\left(\omega + \sqrt{\frac{f_0}{f_2}}\right),
\qquad
\mu_n = \begin{cases}
&0\,, \qquad\qquad\quad\;  \text{for } n \text{ odd}, \\
&\left(\frac{-f_0}{f_2}\right)^{n/2}\,,\quad \;\; \text{for }  n \text{ even}.
\end{cases}
\end{align}
up to prefactor. 

We can calculate the Lanczos coefficients using the Toda-method detailed in~\autoref{app:toda} and, in this particular case, since the spectrum has only a single frequency, the Krylov space is two-dimensional and there are only two Lanczos coefficients~\cite{viswanath1994recursion}.%
\footnote{Every (non-vanishing) frequency determines two Lanczos coefficients. Hence, $L$-frequencies lead to at most a $2L$-dimensional Krylov space~\cite{viswanath1994recursion}.}
As a result, the Krylov exponent is not well-defined, and the UOGH is not applicable.

For a discrete power spectrum with sufficiently many 
distinct, non-zero frequencies, one could hope that the Krylov complexity can experience a window of exponential growth, and the UOGH may again be applicable. Indeed, as we will see in section \ref{sec:discrete}, for shift symmetry constraint equations of high degree $m$, one can generate discrete power spectra determined by an arbitrarily large set of frequencies.

\subsection{Solutions with (anti)-periodic power spectra} \label{sec:(anti)-periodic_power_spectra} 
Motivated by the behaviour observed in eq.~\eqref{eq:lanczos_f1_lambdaf2_asympt}, we now construct an entire family of autocorrelators that solve the shift symmetry constraints and exhibit submaximal Krylov growth. 
For this purpose, it is advantageous to analyse the shift symmetry constraints in frequency space, where the constraint equation becomes the recurrence relation~\eqref{eq:Fdi}.
The solution that we found in eq.~\eqref{eq:perps} is periodic in imaginary frequency, and we will look for solutions with this property. We will find that, to have this periodicity property in its solutions, the shift symmetry constraint equation must be of a particular form, which makes its solutions simple closed-form expressions.

\subsubsection{Conditions for periodicity in the frequency domain}

We start by adopting a quasi-periodic ansatz for the power spectrum
\begin{align}\label{eq:qperi}
	\Phi(z) = C \,\Phi(z+1)\,,\qquad \omega = -i\lambda z\,,
\end{align} 
where, for now, $C$ is an arbitrary real constant.

Plugging the ansatz in eq.~\eqref{eq:qperi} into the recurrence relation capturing the shift symmetry constraint in frequency space in eq.~\eqref{eq:Fdi}, it follows that we must have $C = (-1)^m$ together with the following condition on the roots $\alpha_i$ of polynomial $p(z)$, see eq.~\eqref{eq:factor_polys}:
\bne \label{eq:periodic_freq_zero_str} 
\prod_{i=1}^m (z+1 +\alpha_i ) = \prod_{i=1}^m (z- \alpha_i)\,. \ene
which is equivalent to demanding that 
\begin{align}\label{eq:pzrel} 
     p(-z-1) = (-1)^m p(z)\,. 
\end{align}
So, the roots of $p(z)$ must be reflection symmetric across $-1/2$: $p(\alpha_i) = 0 \Leftrightarrow p(-\alpha_i-1) =0$.

Note that using the definition of the polynomial $p(z)$ in eq.~\eqref{eq:Fdi}, expanding the above relation and matching in powers of $z$ results in the following recursion relation for the functions $f_n(\lambda)$:
\bne \sum_{k=n}^m f_k (\lambda) (-\lambda)^k \begin{pmatrix} k \\ n \end{pmatrix} = (-1)^m f_n (\lambda) \lambda^n \,,
\label{eq:fncon}\ene
which is trivially satisfied when $m=n$. Hence, for any  $f_n$ obeying~\eqref{eq:fncon}, the solution to the shift constraint~\eqref{eq:Fdi} in the frequency domain is (anti)-periodic: periodic for even $m$ and anti-periodic for odd $m$.

\subsubsection{Solving the shift symmetry constraints}\label{sec:anti_per_KMS}

We now turn to the form of the shift symmetry constraint in the time domain given in eq.~\eqref{eq:shifpintro}, 
and specialise to polynomials $p(z)$ obeying eq.~\eqref{eq:pzrel}. In this case, eq.~\eqref{eq:shifpintro} becomes
\begin{align}
    \label{eq:shifp_id} 
p(-\lambda^{-1} \del_t)((-1)^m e^{\lambda t}+1)g_{\hat V   }(t)) = 0\,. 
\end{align}
A given root $\alpha_i$ of the polynomial $p(z)= \lambda^m f_m(\lambda)\prod_{i=1}^m (z - \alpha_i)$
can occur multiple times, but, for convenience, we will here assume that the roots are simple, and defer the general case to \autoref{app:nonsimpleroots}. 
 
The autocorrelators, obtained by solving the shift symmetry constraint given in eq.~\eqref{eq:shifp_id} and performing the imaginary shift $t\rightarrow t+i(\beta/2)$, take the form 
\begin{align}\label{eq:genso2_C}
    C(t) = \frac{ \sum_{q=1}^{m} c_q\, e^{-i\pi \alpha_q} e^{-\alpha_q \lambda t} }{ 1 - (-1)^m e^{\lambda t} }\,,
\end{align}
where we have used that $\lambda=2\pi/\beta$ and that the $\alpha_q$ are all distinct, with at most one equal to zero. Note that \eqref{eq:genso2_C} is in the kernel of the differential operator in eq.~\eqref{eq:shifp_id} because $\alpha_i$ are the roots of $p$: $p(-\lambda^{-1}\del_t) e^{-\alpha_i \lambda t} = p(\alpha_i)e^{-\alpha_i \lambda t} = 0$. These solutions are meromorphic functions with simple poles at $\lambda t =  i \pi (2 n -m)$, with $n \in \mathbbm{Z}$.

To compute the corresponding power spectra, we must distinguish between even and odd values of $m$. 
The power spectra are
\begin{align}\label{eq:Phi_spec_case_noKMS}
\Phi_\mathrm{simp}(\omega) =\begin{cases}
    & \displaystyle   \sum_{q=1}^m c_q\, e^{-i\pi \alpha_q}
\csc\left( \pi \alpha_q - i \frac{\beta \omega}{2} \right)\,,\qquad \text{for } m \text{ odd}\,, \\
   & \displaystyle  \sum_{q=1}^m c_q\, e^{-i\pi \alpha_q}
\cot\left( \pi \alpha_q - i \frac{\beta \omega}{2} \right)\,,\quad\, \text{for } m \text{ even}\,.
\end{cases}
\end{align}
The subscript emphasises that this spectrum corresponds to the case where all roots are simple. As usual, we have not imposed the normalisation condition yet, so the expressions are up to a normalisation factor.

The two-point functions in eq.~\eqref{eq:genso2_C} and associated power spectra in eq.~\eqref{eq:Phi_spec_case_noKMS} encompass a large family of solutions to the shift symmetry constraints. However, in order to arise from consistent thermal systems, these solutions also need to satisfy the KMS condition.
The KMS condition is most straightforwardly imposed on the autocorrelator in eq.~\eqref{eq:genso2_C} by requiring $C(t)=C(-t)$, imposing  the following constraint on coefficients $c_i$ and roots $\alpha_i$:
\begin{align}
    (-1)^m \sum_{q=1}^m c_q e^{2\pi i \alpha_q} e^{(\alpha_q +1)\lambda t} = \sum_{q=1}^m c_q e^{-\alpha_q \lambda t}\,.
\end{align}
For this equality to hold for all times $t$, the roots $\alpha_q$ must come in pairs reflection-symmetric across $-\frac{1}{2}$, consistent with eq.~\eqref{eq:pzrel}, 
\begin{align}\label{eq:zeroes_paired}
\alpha_{j\pm} = -\frac{1}{2} (1\pm \kappa_j)\,,
 \end{align}
 for some $\kappa_i$, 
 and where the associated coefficients are related:
\begin{align}
     c_{j+} = (-1)^{m+1}\, c_{j-}\, e^{-i \pi \kappa_j}\,.
\end{align}
Note that a self-paired root equal to $ -\frac{1}{2}$ is only allowed if $m$ is odd; otherwise, its coefficient must vanish. The pairing of roots in eq.~\eqref{eq:zeroes_paired} mirrors the combinations in eq.~\eqref{eq:antisymm_symm}, when $f_1 = \lambda f_2$, already encountered in the $m=2$ solution.
          
We are now in a position to write down the general expression for the thermal shift-symmetric autocorrelator which satisfies the periodicity condition given in eq.~\eqref{eq:qperi} and its associated power spectrum. Using the relation between the coefficients in eq.~\eqref{eq:genso2_C} arising from imposing the KMS condition, one obtains, by summing over pairs of roots, that 
\begin{align} \label{eq:gencounterexamples}
    C(t) = \begin{cases} 
				&\displaystyle\frac{\sum_{i=1}^{\left \lfloor{m/2}\right \rfloor}  a_i \sinh(\kappa_i \lambda t/2)}{\sinh(\lambda t/2)} \,, \qquad \,\text{even } m \\
			&\displaystyle	\frac{\sum_{i=1}^{\left \lfloor{m/2}\right \rfloor} a_i \cosh(\kappa_i \lambda t/2)}{\cosh(\lambda t/2)}\, , \qquad \text{odd } m \\
			\end{cases}
\end{align}
where the coefficients $a_j$ are arbitrary, up to the normalisation condition. 
Since the roots $\alpha_i$ are reflection-symmetric across $-1/2$, and we assumed that the roots are simple, there is a single root equal to $-1/2$, giving a single $\kappa_i = 0$, if and only if $m$ is odd.

We require that $|\kappa_i| < 1$, so that $C(t)$ decays as $t \to \pm \infty$ and then, using eq.~\eqref{eq:FT_sinh/sinh}, we can perform the Fourier transform, leading to the power spectra  (up to an irrelevant constant prefactor which is determined by the normalisation condition~\eqref{eq:normalisation_Phi}) 
			\begin{align} \label{eq:(anti)periodic_power_spectra}
			\Phi(\omega) = \begin{cases} 
				&\displaystyle\sum_{i=1}^{\left \lfloor{m/2}\right \rfloor} \frac{a_i \sin (\pi \kappa_i)}{\cos(\pi \kappa_i) + \cosh(\beta \omega)}\,, \qquad \qquad\text{even } m \\
				&\displaystyle\sum_{i=1}^{\left \lfloor{m/2}\right \rfloor} \frac{a_i \cos(\pi \kappa_i /2) \cosh(\beta \omega/2)}{\cos(\pi \kappa_i) + \cosh(\beta \omega)} \,, \qquad \text{odd } m\,,
			\end{cases}
			\end{align}
 where in the odd case, the unpaired zero $\alpha_j=-1/2$, is accounted for by having a term with $\kappa_j=0$. Note that $\Phi(\omega - i\lambda) = (-1)^m \Phi(\omega)$, where again $\lambda = \frac{2\pi}{\beta}$. Finally, let us point out that the spectra are invariant under $\lambda\to-\lambda$, which is the final requirement as explained below eq.~\eqref{eq:Fdi}.

For large $\omega$, the power spectra have the asymptotes
\begin{align} \label{eq:largeomegaevenodd}
            \Phi(\omega)\sim    \begin{cases} &
  e^{-\beta \omega} \sum_{i=1}^{\left \lfloor{m/2}\right \rfloor} a_i \sin(\pi \kappa_i)\,, \qquad \quad\text{even } m \\
 &e^{-\beta \omega / 2} \sum_{i=1}^{\left \lfloor{m/2}\right \rfloor} a_i \cos(\pi \kappa_i/2)\,, \quad\; \text{odd } m
\end{cases}\quad\; \text{ as }\,  \omega \to +\infty\,.
\end{align}
That is, unless the weighted sum conspires to vanish. For even $m$, $\Phi(\omega)$ decays as in eq.~\eqref{eq:decay_Phi_gen}, taking $\rho=2$, at a rate $\omega_0 = T$. For odd $m$, the rate equals to $\omega_0 = 2T$. We conclude that the odd case leads to the expected value for maximal chaos, while the even case leads to half-maximal exponents%
\footnote{This assumes that the Lanczos coefficients are asymptotically linear, so that $\lambda_K = \pi \omega_0$, which we have only explicitly shown for $m=2$, in eq.~\eqref{eq:lanczos_f1_lambdaf2}.}. 
We will return to this observation and its possible implications for the bound~\eqref{eq:conjecture} in \autoref{sec:interpr}. 

\paragraph{Shift symmetry constraint on the power spectrum exponent. } The decay of the power spectrum can be directly constrained from shift symmetry constraints. To this end, consider the $\omega \to \infty$ asymptote of the shift symmetry constraint in the frequency domain, see eq.~\eqref{eq:fts}. If, in the time domain, the shift symmetry constraint is an $m$-th order ODE, then, by taking the leading term in the frequency domain, we get the asymptotic (anti)-periodicity condition 
\bne \Phi(\omega \mp i\lambda) \sim (-1)^m \Phi(\omega), \qquad \omega \to \infty. \label{eq:aspos} \ene 
This is precisely the same as the ansatz for $\Phi(\omega)$ that we assumed in~\eqref{eq:qperi}, with $C = (-1)^m$. Importantly, this constraint arises as a consequence of the shift symmetry constraint and should only hold asymptotically. 

A corollary: if we assume that the power spectrum asymptotically decays exponentially, $\Phi(\omega) \sim e^{ - \omega / \omega_0 }$  as $\omega \to \infty$, then~\eqref{eq:aspos} implies the following constraint on the decay exponent:
\begin{align} \label{eq:omega0constraint}
\frac{1}{\omega_0 \beta} \in
\begin{cases}
\;\mathbb{Z}\,, & m \text{ even} \\
\;\mathbb{Z} + \tfrac{1}{2}\,, & m \text{ odd}
\end{cases}
\end{align}

This is indeed consistent with~\eqref{eq:largeomegaevenodd}, where we saw that $\omega_0 = \pi T$ for even $m$ and $2\pi T$ for odd $m$. 

\subsubsection{Simple examples}

Let us now turn to several concrete examples of polynomials $p(z)$ verifying the constraint in eq.~\eqref{eq:periodic_freq_zero_str} arising from demanding periodicity in frequency space and the corresponding Lanczos coefficients. 

\paragraph{$\bm{m=2}$ and Lanczos coefficients.} For $m = 2$, the only solution to the periodicity condition in terms of the polynomial $p(z)$ in eq.~\eqref{eq:pzrel} is $f_1 = \lambda f_2$, for which the pair of roots are
\bne \alpha_\pm = \frac{1}{2}\left(-1\pm \sqrt{1-\frac{4f_0}{f_2 \lambda^2}}\right). \ene
One can verify that substituting this expression in the expression for the autocorrelation in eq.~\eqref{eq:genso2_C} yields a two-point function $g_{\hat V}(t)$ that coincides with the solution obtained by directly solving the shift symmetry constraints under the identification $f_1 = \lambda f_2$. The Lanczos coefficients are given in eq.~\eqref{eq:lanczos_f1_lambdaf2}, with asymptotes in eq.~\eqref{eq:lanczos_f1_lambdaf2_asympt}.

\paragraph{$\bm{m=3}$ and Lanczos coefficients.} For $m =3$, in order to satisfy the condition on the polynomial $p(z)$ in eq.~\eqref{eq:pzrel}, we require $f_0 = \frac{\lambda}{2}f_1 - \frac{\lambda^3}{4}f_3$ and $f_2 = \frac{3}{2} \lambda f_3$, for which the three roots of $p(z)$ are
\bne \alpha_1 = -\frac{1}{2}, \quad \alpha_\pm =-  \frac{1}{2}\left(1\pm \sqrt{3-\frac{4 f_1}{\lambda^2f_3}}\right). \label{eq:m3ex} \ene
To verify the KMS condition, we must have that $c_1 = 0$, while $c_\pm$ can be non-zero and the ``offset'' parameter defined in eq.~\eqref{eq:zeroes_paired} is $\kappa = (3-4(f_1/\lambda^2f_3))^{1/2}$
The associated KMS-symmetric correlator, restricting to values $0<\kappa<1$ to ensure convergence of the Fourier transform to the associated power spectrum, is 
\bne C(t) =  \frac{\cosh(\kappa \lambda t/2)}{\cosh(\lambda t/2)}\,,\ene
from which the Lanczos coefficients can be readily obtained from the method reviewed in appendix \ref{app:toda}. The exact expression for the coefficients is
\bne b_n^2 = \frac{\lambda^2}{4} \left((n+1)^2 - \frac{1+(-1)^n}{2}\kappa^2\right) \,.\ene
Identifying, at large $n$, the  $b_n\sim  (\lambda_K/2) n$, this leads to the Krylov exponent $\lambda_K =2\pi T$.

\subsection{Even and odd-order shift symmetry constraints}\label{sec:discrete}
We now turn to the second class of special cases identified in eq.~\eqref{eq:f10}: the time-periodic autocorrelation function, or equivalently, the discrete power spectra. They arise from the shift symmetry constraints formulated in eq.~\eqref{eq:shift} when the odd or even terms vanish. Without loss of generality, consider the case with only the even terms\footnote{For a shift symmetry constraint equation that is odd-order, we can similarly simplify it to
\begin{align}
\partial_t\left(\sum_{n=0}^{\frac{m-1}{2}}f_{2n+1}(\pm\lambda)\partial_t^{2n}C(t)\right)=0\,.
\end{align}
The KMS-invariant solution to this equation is the same as~\eqref{eq:autocorr_discrete} up to the addition of a constant.
}
\begin{align}\label{eq:shift_diff_even}
	\sum_{n=0}^{m}f_{2n}(\pm\lambda)\partial_t^{2n}C(t)=0\,.
\end{align}
This equation can be solved by
\begin{align}\label{eq:autocorr_discrete}
	C(t)=\sum_\ell c_{2,\ell} \cos ( \omega_\ell t)\,,
\end{align}
which is KMS symmetric. The sum in $\ell$ runs over the number of solutions to $0=\sum_{n=0}^m f_{2n}(\pm\lambda) \omega_\ell^{2n}$ which determines the frequencies $\omega_\ell$. 
The Hermiticity condition demands that the correlator is real and even, implying that $c_{2,\ell} \in \mathbb{R}$. 

In chaotic theories at finite temperature, we do not expect oscillating solutions like eq.~\eqref{eq:autocorr_discrete}. Instead, we expect thermal decay, see~\autoref{app:phys_props}. So, we could rule out these solutions on physical grounds, but we will continue regardless to better understand the implications of such non-thermal behaviour.

The power spectrum is readily obtained by the Fourier transformation of eq.~\eqref{eq:autocorr_discrete}:
\begin{align}\label{eq:shift_sol_discrete}
	\Phi(\omega) =\pi \sum_{\ell=0}^{m}  c_{2,\ell} \left( \delta(\omega - \omega_\ell ) + \delta(\omega + \omega_\ell ) \right)\,.
\end{align}
Note that similar discrete power spectra were also considered in~\cite{Anegawa:2024wov, Anegawa:2024yia}.

Away from the continuum limit, the moments can be easily computed from eq.~\eqref{eq:moments} to be
 \begin{align}
     \mu_n = \begin{cases}
&0\,, \qquad \qquad \qquad \qquad \; \; n \text{ odd}\,, \\
&\displaystyle \sum\limits_{\ell = 0}^m 
c_{2,\ell} \,  \omega_\ell^n\,,\quad\;\;   n \text{ even}\,.
\end{cases}
 \end{align}
For a sufficiently large set of frequencies $\omega_\ell$, while the Krylov space is finite dimensional, the resulting Lanczos coefficients may show a non-smooth behaviour before the sequence of coefficients ends. The UOGH assumes a continuous power spectrum and thus does not directly apply to systems with a purely discrete spectrum, where the spectral measure consists of isolated eigenvalues. It may, however, remain approximately valid if the eigenvalues are sufficiently close to a dense set, effectively approximating a continuous spectrum. Nevertheless, we emphasise that these power spectra still satisfy the shift symmetry constraints. This ensures that, within the time and temperature range for which the EFT is an accurate description of the microscopic theory, the system realises maximal chaos in the sense of having a maximal Lyapunov exponent. 

\subsection{Interpretation of the submaximal Krylov exponents within the EFT}\label{sec:interpr}

\begingroup
\setlength{\tabcolsep}{6pt} 
\renewcommand{\arraystretch}{2} 
\begin{table}[t]
\small
\centering
\begin{tabular}{p{5cm}|c|c|c}
 & $C(t) $ & $\Phi(\omega) $ & Krylov exponent  \\ \hline \hline
Schwarzian correlator (sec.~\ref{sec:Schwarzian}) &   $\cosh(\pi t\beta)^{-2\Delta}$   &  $\left|\Gamma\left(\Delta+ \frac{i\omega\beta }{2 \pi }\right)\right|^2$            &   Maximal  $\lambda_K=\lambda_L$           \\ \hline
Power spectrum periodic in imaginary frequency (sec.~\ref{sec:(anti)-periodic_power_spectra}) &    $ \sum_i\frac{\sinh(\kappa_i \lambda t/2)}{\sinh(\lambda t/2)}$    &   $\sum_i \frac{1}{\cos(\pi \kappa_i) + \cosh(\beta \omega)}$             &      Half-maximal  $\lambda_K=\lambda_L/2$                   \\ \hline
Power spectrum anti-periodic in imaginary frequency (sec.~\ref{sec:(anti)-periodic_power_spectra})  &     $ \sum_i\frac{\cosh(\kappa_i \lambda t/2)}{\cosh(\lambda t/2)}$     &  $\sum_i \frac{\cosh(\beta \omega/2)}{\cos(\pi \kappa_i) + \cosh(\beta \omega)}$ &         Maximal   $\lambda_K=\lambda_L$              \\ \hline
Discrete~power~spectrum $\qquad$ (sec.~\ref{sec:discrete}) &  $\sum_\ell \cos(\omega_\ell t)$      &    $\sum_\ell  \delta(\omega -\omega_\ell)$     &       Finite set of staggering $b_n$        
\end{tabular}
\caption{Summary of the solutions to the shift symmetry constraints obtained in section \ref{sec:Krylov_vs_EFT}. To keep the table concise, expressions are presented schematically; detailed formulae can be found in the main text. 
Notably, all correlator solutions to the even-order $m$ shift symmetry constraint equation that are periodic in imaginary frequency have a half-maximal Krylov exponent.
}\label{table:classes_sols}
\end{table}
\endgroup

In the previous sections, we considered several examples of two-point functions that satisfy the shift symmetry constraints, as well as a minimal set of physical constraints, see \autoref{table:classes_sols}. Within this space of solutions, a subset satisfying a periodicity condition in imaginary frequency was simple enough that we could calculate the Lanczos coefficients analytically and find $\lambda_K=\pi T$.
This result indicates that within the EFT of maximal chaos, the shift symmetry guarantees maximal Lyapunov growth in the OTOC diagnostic as defined in the EFT framework, while it does not uniquely determine the Krylov exponent extracted from thermal two-point functions. Interpreting this in terms of the conjectured bound $\lambda_{L} \leq \lambda_K \leq 2\pi T$ requires care, since at finite temperature the exponents depend on the choice of regularization scheme. The apparent tension arises when one adopts a broad interpretation of the conjecture, in which $\lambda_L$  is identified with the Lyapunov exponent extracted from the OTOC as defined in \cite{Blake:2017ris,Blake:2021wqj}, and/or when one assumes that maximal chaos as seen through the OTOC is a physical statement and thus independent of this choice. Let us highlight that for maximally chaotic systems such as those described by the EFT construction, $\lambda_L$ is expected to be universal ($2\pi T$) and robust against changes in regularization. See \autoref{footnote4} for references and discussion. To critically assess these results, it is also important to contrast the regime of validity and assumptions of the EFT of maximal chaos and the UOGH.

\begin{itemize}
     \item {\bf Infinite vs finite \boldmath{$N$}.}  The Krylov exponents obtained in this work arise from the autocorrelator of the bare field, which corresponds to the leading-order contribution in the $1/N$ expansion to the autocorrelator of the full field; see eq.~\eqref{eq:full_to_bare_g}. This level of approximation is sufficient to reproduce the maximal Lyapunov exponent in the OTOC, and as demonstrated in \cite{Blake:2017ris,Blake:2021wqj}, where the addition of the shift symmetry constraints ensures the expected exponential growth of OTOCs governed by the Lyapunov exponent. However, we find that in the leading-order truncation, some correlation functions yield half-maximal Krylov exponents.

This leading-order feature is most extreme in the oscillatory solution of eq.~\eqref{eq:oscillatory_sol}. Here, the operator is effectively confined to a finite-dimensional Krylov subspace at leading order, yet the OTOC remains maximally chaotic. This apparent paradox is due to the structure of the large-$N$ expansion. The maximal Lyapunov growth in the OTOC arises from the exchange of the hydrodynamic mode, which is a $1/N$ effect. In contrast, the operator growth diagnosed by the leading-order $O(1)$ contribution to the correlation function remains blind to the chaotic nature of the EFT mediated by the hydrodynamic mode.

This raises a natural question: could the inclusion and resummation of the full tower of $1/N$ corrections, particularly those associated with the cloud field, restore maximality and reconcile the discrepancy? Notably, the Lyapunov exponent is defined from a large $N$ expansion of the OTOC, while the Krylov exponent does not need an expansion of the Krylov complexity to be defined.
Therefore, the Krylov exponent has been proposed as a non-semiclassical diagnostic of chaos, in contrast to the Lyapunov exponent. 
Nevertheless, barring special fine-tuning or the presence of additional symmetries, it seems highly unlikely that resumming these corrections would modify the leading-order result by a factor of two.

Finally, let us reiterate that the generalised bound\footnote{This refers to the bound even when assuming monotonicity of the Lanczos coefficients, whether formulated in terms of the decay rate of the power spectrum or the slope of the Lanczos coefficients themselves.} in eq.~\eqref{eq:Kchabintro} remains conjectural. Partial evidence in support of this conjecture has been presented, notably by \cite{Avdoshkin:2019trj}. More recently, \cite{Gu:2021xaj} showed that the bound holds in a broad class of theories. However, the derivation in \cite{Gu:2021xaj} crucially relies on a precise factorisation structure of the OTOC at all orders in the $1/N$ expansion. While the EFT framework employed here does exhibit such factorisation at leading order in $1/N$, it imposes no constraints on higher-order corrections, and moreover, there are technical obstacles to imposing factorisation at higher order in $1/N$, leaving the validity of the provided argument in this context an open question.

\item {\bf IR vs. UV degrees of freedom.} 
An important question is what mechanism ensures $\lambda_K = 2\pi T$ in maximally chaotic systems. One proposed route is via the analytic structure of the two-point function, which is controlled, for example, by conformal symmetry or a UV regulator. For instance, in QFTs, the thermal Wightman correlator defines a well-regulated inner product for local operators, and in 2D CFTs on a plane, it takes a universal form fixed entirely by symmetry. Local continuum QFTs have poles in $C(t)$ when the operators are coincident on the thermal circle, at $t = \pm i\beta/2$, and when the Lanczos coefficients are smooth 
functions of $n$, this leads to $\lambda_K = 2\pi T$~\cite{Dymarsky:2021bjq}. This behaviour appears even in integrable or free theories, where the exponential growth of Krylov complexity reflects the correlator’s analytic structure rather than genuine chaos. We refer the reader to \cite{Rabinovici:2025otw} for a recent comprehensive discussion. 

To isolate chaos-sensitive contributions, some have proposed adding a UV cutoff or discretising the theory on a lattice \cite{Camargo:2022rnt,Dymarsky:2021bjq,Avdoshkin:2022xuw}. In our case, the EFT naturally includes a UV cutoff since all the UV degrees of freedom are integrated out. Yet we find that the Krylov bound $\lambda_K = 2\pi T$ is not automatically realised.\footnote{Note that if one naively takes the infinite temperature limit in these examples, an apparent contradiction arises with a proven bound \cite{Parker:2018yvk} stating that, at infinite temperature, the Krylov exponent must bound the Lyapunov exponent from above. However, as the temperature approaches infinity, the IR/UV scale separation required to define the local effective action in the EFT breaks down because the length scale of thermal fluctuations is smaller than the UV cutoff scale. Consequently, the finite-temperature EFT examples considered here do not straightforwardly extend to an infinite-temperature theory, which necessarily involves all degrees of freedom beyond the EFT description.} While the shift symmetry is sufficient to ensure maximal Lyapunov growth, we have found that it is not sufficient to ensure maximal Krylov exponent.

This raises a conceptual puzzle. Since the exponential growth in time of Krylov complexity unfolds within the relaxation-to-scrambling time window where EFT is expected to be an accurate approximation of the microscopic theory, one might ask if UV degrees of freedom still play a role despite the EFT reliably capturing $C(t)$. A clue lies in the definition of Krylov complexity. While $C(t)$ is well-described by the EFT over the aforementioned timescales, its short-time expansion involves a tower of moments $\mu_{n}$, see eq.~\eqref{eq:moments}, which are increasingly sensitive to tails in the spectrum. The Lanczos coefficients $b_n$ are recursively built from these moments, so their asymptotic growth, and hence the Krylov exponent, can reflect physics beyond the regime where the EFT is a reliable description. In other words, the moments are sensitive to UV details and hence may lie outside the scope of the physics captured by the EFT. This emphasises that while the EFT provides a controlled setting to study chaos through the OTOC, the full Krylov dynamics may encode global spectral information that is only visible in the microscopic theory. Consequently, our results suggest that the success of a given UV regulator in enforcing $\lambda_K = 2\pi T$ depends sensitively on how it interacts with these UV-sensitive structures.
\end{itemize}

\section{Solving the shift symmetry constraints in full generality} \label{sec:solv_shift}

In this section, we shift our focus away from Krylov complexity to an analysis exploring the solutions to the shift symmetry constraints and the thermal systems they describe\footnote{Note that we will also no longer normalise the power spectra and the autocorrelator as prescribed in \autoref{sec:shift_time_freq_C_phi}.}. In particular, we first turn to the symmetry constraints in eq.~\eqref{eq:Fdi} in the frequency domain, where it is a recurrence relation, which we solve. This leads to a general power spectrum solution for the shift symmetry constraints. We then transition to the time domain, and give the general solution there by transforming the shift symmetry constraints in eq.~\eqref{eq:shift} to the family of differential equations that are solved by Meijer G-functions.

\subsection{General solution in the frequency domain}\label{sec:gen_sol_freq_space}
For convenience, let us restate the first-order non-homogeneous linear recurrence relation in eq.~\eqref{eq:Fdi}, which is the shift symmetry constraint in the frequency domain\footnote{Recall that we have a second shift symmetry constraint, and for the power spectrum $\Phi(\omega)$ to satisfy both shift symmetry constraints simultaneously, it should be invariant under $\lambda\to-\lambda$ which we impose at the end of the analysis.} 
\begin{align}\label{eq:Fdi_r}  
\Phi(z+1) = \frac{p(z)}{p(-z-1)} \Phi(z)\,,\qquad p(z) = \sum_{n=0}^m f_n(\lambda) (\lambda z)^n\,,
\end{align} 
where we defined $\omega = -i\lambda z$.
The general solution to this equation is given by
\bne \Phi(z) = \tilde{\Phi}(z) G(z)\,, \ene
where  $\tilde{\Phi} (z)$ is a particular solution to eq.~\eqref{eq:Fdi} and $G(z)$ is the homogeneous solution satisfying  $G(z+1) =  G(z) $.
Since $p(z)$ is a polynomial, it can be expressed as a product of its $m$ roots $\alpha_i$, as in eq.~\eqref{eq:factor_polys}, and thus
\bne \frac{p(z)}{p(-z-1)} = (-1)^m \prod_{i=1}^m \frac{z-\alpha_i}{z+ 1+\alpha_i}\,. \ene
We aim to find a particular solution $\tilde{\Phi}$. To this end, we start from the ansatz $\tilde{\Phi}(z) = e^{im\pi z} \prod_{i=1}^m \tilde{\Phi}_i (z)$.
This ansatz is a solution to eq.~\eqref{eq:Fdi} if each factor satisfies
\bne \tilde{\Phi}_i (z+1) =  \frac{z-\alpha_i}{z+1+\alpha_i} \tilde{\Phi}_i(z)\,, \ene
which is solved by $\tilde{\Phi}_i(z) = \frac{\Gamma(z-\alpha_i)}{\Gamma(z+1+\alpha_i)}$.
Thus the most general solution to eq.~\eqref{eq:Fdi} is
\begin{align}\label{eq:solsPhiuniv}
     \Phi(z) =   e^{im \pi z} \left( \prod_{i=1}^m \frac{\Gamma(z-\alpha_i)}{\Gamma(z+1+\alpha_i)}\right ) G(z) \,, \quad \text{ where}\quad G(z) = G(z+1)\,.
\end{align}
So, the general solution to the shift symmetry constraints is fixed by the roots $\alpha_i$ of the polynomial $p(z)$ constructed from the functions $f_n(\lambda)$, up to an arbitrary periodic function $G(z)$. We will return to $G(z)$ momentarily.

Returning to the original $\omega$ variable, and using that we can absorb or pull any periodic function from $G(z)$, we can rewrite eq.~\eqref{eq:solsPhiuniv} as
\begin{equation}\label{eq:solsPhi}
\begin{split}
    \Phi(\omega) &=\begin{cases}
        &\dfrac{1}{\tanh\left(\frac{\pi \omega}{\lambda}\right)}  \displaystyle\prod_{j=1}^m \frac{\Gamma(\frac{i\omega}{\lambda} - \alpha_j)}{\Gamma(\frac{i\omega}{\lambda} +1+ \alpha_j)} G(\omega)\,,\quad m \, \rm even\\
        &\dfrac{1}{\sinh\left(\frac{\pi \omega}{\lambda}\right)}  \displaystyle\prod_{j=1}^m \frac{\Gamma(\frac{i\omega}{\lambda} - \alpha_j)}{\Gamma(\frac{i\omega}{\lambda} +1+ \alpha_j)} G(\omega)\,,\quad  m \, \rm odd
    \end{cases}
\end{split}
\end{equation}
where in each case $ G(\omega - i\lambda) = G(\omega)$.
We used that $e^{\frac{-m\pi \omega}{\lambda}}/\tanh(\frac{\pi\omega}{\lambda})$ is periodic in $\omega$ with period $i\lambda$ when $m$ is even and that $e^{\frac{-m\pi \omega}{\lambda}}/\sinh(\frac{\pi\omega}{\lambda})$ is periodic in $\omega$ with period $i\lambda$ when $m$ is odd. 

Note in passing that, if we set $G=1$, while the odd case automatically displays asymptotic exponential decay, the even case does not  
\begin{align} \label{eq:g1asymp}
\Phi(\omega) \sim
\begin{cases}
& \omega^{-2\sum_j \alpha_j - m}\,, \qquad \qquad \;\;\; m \text{ even} \\
& e^{-  \pi \omega / \lambda} \, \omega^{-2\sum_j \alpha_j - m} \,, \qquad m \text{ odd}
\end{cases}\quad
\quad \text{as } \omega \to \infty\,,
\end{align}
because in the odd case, the exponent arises from $\sinh(\pi\omega/\lambda)^{-1}$ factor, while in the even case, the $\tanh(\pi\omega/\lambda)^{-1}$ factor asymptotes to 1. Note that the parameters $\alpha_j$ do not affect the exponential decay, only entering as polynomial powers.

A thermal system at finite temperature must have a power spectrum that decays at least exponentially at large frequencies, see \autoref{app:phys_props}. Pure power-law decay violates this, as it would imply that the real-time correlator is not analytic in any strip around the real axis.
As can be seen from eq.~\eqref{eq:g1asymp}, this physical condition rules out $G = 1$ for even $m$.

Examples of periodic functions that $G(\omega)$ can include are a constant, a combination of Jacobi functions, or ratios of trigonometric functions. 
To build some intuition, let us first consider a simple example. 

\paragraph{Example: conformal correlator.} Consider the shift symmetry constraint for the case $f_0 = \lambda \Delta$ and $f_1 = 1$. In this case, the polynomial $p(z)$ is simply $p(z)=\lambda(z+\Delta)$.
Hence, it has only one root: $\alpha_1 = -\Delta$. Plugging this root into the general solution~\eqref{eq:solsPhi} gives the power spectrum
\begin{align}\label{eq:power_sp_schwarzian}
	\Phi(\omega) =   \frac{1}{\sinh\left(\frac{\pi \omega}{\lambda}\right)} \frac{\Gamma(\frac{i\omega}{\lambda} +\Delta)}{\Gamma(\frac{i\omega}{\lambda} +1 - \Delta)} G(\omega)\,,
\end{align}
where $G(\omega)$ is any function that has period $i\lambda$. Of course, this solution has to satisfy the KMS condition, which translates to imposing $\Phi(\omega)=\Phi(-\omega)$. Hence, $G(\omega)$ additionally satisfies
\begin{equation}
    \frac{G(\omega ) \Gamma \left(\Delta +\frac{i \omega }{\lambda }\right)}{\Gamma \left(\frac{i \omega }{\lambda }+1-\Delta\right)}+\frac{G(-\omega ) \Gamma \left(\Delta -\frac{i \omega }{\lambda }\right)}{\Gamma \left( -\frac{i \omega }{\lambda }+1-\Delta\right)}=0\,.
\end{equation}
A solution compatible with these constraints is the following choice of $G(\omega)$
\begin{align}\label{eq:G_Schwarzian} 
G(\omega) =\frac{\Gamma(\Delta)^2}{\pi} \frac{\sinh\left(\frac{\pi \omega}{\lambda}\right)}{\sin \left(\pi \left(\Delta - \frac{i\omega}{\lambda}\right)\right)} \,,
\end{align}
For this choice, the power spectrum reduces to
\begin{align}\label{eq:app_SYK_power_spectrum}
\Phi(\omega) =  \frac{ 1 }{ \Gamma(\Delta)^2}
\Gamma \left(\Delta -\frac{i \omega }{\lambda }\right) \Gamma \left(\Delta +\frac{i \omega }{\lambda }\right)\,,
\end{align}
where we have imposed the conventional normalisation such that $\Phi(0)=1$.
Inverse Fourier transforming, or Mellin-Barnes transforming, to the time domain 
\begin{align}\label{eq:schw_MB}
\cosh^{-2\Delta}\left( \frac{\pi t}{\beta} \right)
=\int_{-\infty}^{\infty}\mathrm  d\omega\,
2^{2\Delta } \frac{\beta}{2\pi}
 \frac{1}{\Gamma(2\Delta)}
 \left| \Gamma\left( \Delta + \frac{i\omega\beta}{2\pi} \right) \right|^2
 \frac{e^{i\omega t}}{2\pi} \,,
\end{align}
and applying the usual complex shift leads to, up to a normalisation factor, the following two-point function 
\begin{align}\label{eq:conf_corr}
 g_{\hat V}(t) =  \sinh\left(\lambda t/2  \right) ^{-2\Delta}\,,
\end{align}
matching, as expected \cite{Blake:2017ris}, the  conformal correlator.

\paragraph{Discrete power spectra.}
We can also show how the discrete power spectra of~\autoref{sec:discrete} fit into the general solution given in this section. Of the shift symmetry constraint equations that give discrete spectra, let us focus on the even-order one, which has $p(z) = p(-z)$, for which the shift symmetry constraint~\eqref{eq:Fdi_r} simplifies to
\bne \Phi(z) p(z) = \Phi(z+1) p(z+1)\,, \ene
whose solution is 
\begin{align}
    p(z) \Phi(z) = G(z)\,, \label{eq:discGen}
\end{align}
for any $G(z) = G(z+1)$. The discrete power spectra in eq.~\eqref{eq:shift_sol_discrete} are the solutions for $\Phi(z)$ to eq.~\eqref{eq:discGen} with $G = 0$, with the positions of the delta-functions at the roots of $p(z)$.

\paragraph{Solutions in terms of Meijer's G-functions.}
The symmetry constraints can be solved exactly in terms of Meijer's G-functions. More specifically, the shift symmetry constraints can be mapped systematically to a particular subset of differential equations defining  Meijer's G-function. Specifically, it leads to two main classes
    \begin{align}
        G^{2m,\,0}_{0,\,2m}\!\left(-e^{\lambda t}\left|\begin{array}{c}
-\\
\pm\alpha_{1},\dots,\pm\alpha_{m}
\end{array}\right.\right)\quad \text{and}\quad G^{m,1}_{m,m} \left( -e^{\lambda t} \,\Bigg|\, \begin{array}{c} 0,\ 1 + \alpha_1,\dots, 1 + \alpha_m \\ 1,\ -\alpha_1,\dots, -\alpha_m \end{array} \right)\,,\end{align}
    where the symmetric structure of the parameters results from imposing KMS and reality.  This representation in terms of Meijer's functions provides, at first sight, an analytic continuation and access to systematic asymptotic expansions. This analytic structure is made explicit in \autoref{App:Meijers_hypergeom}, where we construct the Meijer representations. However, it is important to emphasise that the effective theory of maximal chaos is only expected to be an accurate approximation of the microscopic theory within a finite temporal window: not at very early times before equilibration, nor at parametrically late times where scrambling effects dominate. Consequently, while the Meijer function encodes a full analytic continuation of the formal solution, its physical relevance to the microscopic theory is restricted to the intermediate time regime where the effective description is an accurate approximation. 

\subsection{KMS-symmetric solutions}
From the example of the conformal correlator discussed above, it is clear that the KMS condition $\Phi(\omega)=\Phi(-\omega)$ significantly constrains the solution in eq.~\eqref{eq:solsPhi}. In addition, we impose the power spectrum to be real valued, warranting a real spectrum. To illustrate the implications of these conditions on the general solution, let us, for concreteness, consider odd values of $m$, i.e. the second line in eq.~\eqref{eq:solsPhi}. The analysis presented below can be readily carried out for even values of $m$. 

The general solution in the frequency domain obtained in eq.~\eqref{eq:solsPhi} represents a large class of solutions, even for a fixed set of roots $\alpha_i$, since one can choose any periodic function $G$, as long as the resulting power spectrum obeys the physical constraints and is Fourier-invertible. We will focus on three particular choices for $G$ and see, for one of these choices, that once the physical conditions are imposed, a familiar structure emerges: the power spectrum is closely related to the set of quasi-normal modes (QNM) predicted from the thermal product formula of \cite{Dodelson:2023vrw}.

\paragraph{Case 1.}
First, let us consider $G$ to be a constant. We choose this constant to be $\pm i$ to warrant realness; the sign choice will later ensure the positivity of the power spectrum. We find that for this case
\begin{align}\label{eq:nonsimpl_case1}
\Phi(\omega)
= \dfrac{i\sigma}{\sinh\left(\frac{\pi \omega}{\lambda}\right)}  \displaystyle\prod_{j=1}^m \frac{\Gamma(\frac{i\omega}{\lambda} - \alpha_j)}{\Gamma(\frac{i\omega}{\lambda} +1+ \alpha_j)}\,,\qquad \sigma=\pm\,.
\end{align}
A choice of roots for which this function is even (KMS) and real for real $\omega$ is 
\begin{align}\label{eq:set_case1}
    \{\alpha_j\}_{j=1}^m=\Big\{ \frac{a_k}{2} + ib_k,\; \frac{a_k}{2} - ib_k\Big\}_{j=1}^{\ell}
\cup \{c_k\}_{k=1}^{m-2\ell}\,, 
\end{align}
where $a_k \in 2\mathbb{Z}+1\,, \;	b_k \in \mathbb{R}\,, \;	c_k \in \mathbb{Z}\,,$ and $\ell\in[0,1,\ldots,m]$.

Note that whenever a root is a negative integer, $\alpha_j=-n$, the fraction of Gamma functions related to this root becomes a polynomial with a power of $i$ as follows
\begin{align}
     \frac{\Gamma\left(\frac{i\omega}{\lambda} + n\right)}{\Gamma\left(\frac{i\omega}{\lambda} + 1 - n\right)}
=\frac{\lambda}{i\omega} \prod_{k=0}^{n-1} \left(\frac{i\omega}{\lambda} + k\right)\,\left(\frac{i\omega}{\lambda} - k\right)= -\frac{\lambda}{\omega} (i)^{2n+1}\prod_{k=0}^{n-1} \left(\frac{\omega^2}{\lambda^{2}} + k^2 \right)\,.
\end{align}
This yields an overall prefactor of $(-1)^{\sum_{k=1}^{k_{neg}} (n_k-1/2)}$ where the set ${n_k}$ are the absolute values of the negative integer roots and $\ell$ is the number of remaining conjugate pairs. Hence, to make sure that the power spectrum is positive for all $\omega$, we sometimes need an overall $(-1)$ explaining the factor $\sigma$.

Finally, let us analyse how the power spectrum~\eqref{eq:nonsimpl_case1} behaves around $\omega=0$. Expanding around $\omega=0$, using that $\Gamma(\omega-n) \sim (-1)^n/(n!\omega)$ as $\omega \to 0$ for $n\in \mathbb Z_{\geq 0}$, gives
\begin{align}
\Phi(\omega)\sim \omega^{k_\mathrm{neg}-k_{\mathrm{pos},0}-1}, \qquad \omega \to 0\,.
\end{align}
up to a proportionality constant, and where $k_\mathrm{pos,0}$ is the number of zero or positive integer roots in the set~\eqref{eq:set_case1}, and $k_\mathrm{neg}$ is the number of negative integer roots. So, $\Phi(\omega)$ will diverge as $\omega \to 0$ if $k_\mathrm{neg} - k_{\mathrm{pos},0}- 1 < 0$, and remain finite otherwise, with it vanishing if $k_\mathrm{neg} - k_{\mathrm{pos},0}- 1 > 0$.

Another class of sets of roots that give even and real $\Phi(\omega)$ is
 \begin{align}
     \{\alpha_n\}_{n=1}^m=\{0\}\cup\{\gamma_k,-1-\gamma_k, \gamma^*_k,-1- \gamma^*_k\}_{k=1}^\ell\,, \quad \gamma_n\in\mathbb C\,,
 \end{align}
and sets of roots in this class have the additional interesting property that they all give the same power spectrum: 
\begin{align}\label{eq:power_spectrum_totalv2}
\Phi(\omega) = \frac{\lambda}{ \omega \,\sinh\left( \frac{\pi\omega}{\lambda} \right)}\,,
 \end{align}
regardless of the choice of $\gamma_k$.

\paragraph{Case 2.}
Now, let us consider the periodic function
\begin{align}
	G(z)= 
\prod_{j=1}^m	
\frac{i \sin(\frac{i\pi \omega}{\lambda})}
{\sin(\pi (\frac{i \omega}{\lambda}+\alpha_j))}\,,
\end{align}
such that the total solution now reads
\begin{align}
	\Phi(\omega)&= \frac{1}{\sinh\left(\frac{\pi \omega}{\lambda} \right)}
\prod_{j=1}^m	
\frac{\Gamma\big(\frac{i\omega}{\lambda}-\alpha_j\big)}
{\Gamma\big(\frac{i\omega}{\lambda}+1+\alpha_j\big)}
\frac{i \sin(\pi \frac{i\omega}{\lambda})}
{\sin(\pi (\frac{i\omega}{\lambda}+\alpha_j))}\,.
\end{align}
Applying the  Weierstrass-type product formula\footnote{We use the following identity which follows from the  Weierstrass factorisation theorem: \begin{align}\frac{1}{\Gamma(s-z)\Gamma(s+z)} = \frac{1}{\Gamma(s)^2}\prod_{n=0}^\infty \left( 1 - \left(\frac{z}{n+s} \right)^2 \right).
\end{align}}, allows us to further simplify this expression and find 
\begin{align}
	\Phi(\omega)
&= \pi^{-m}	\left(\sinh\left(\frac{\pi \omega}{\lambda}\right)\right)^{m-1}\prod_{j=1}^m	
\Gamma\left(\frac{i\omega}{\lambda}-\alpha_j\right)
\Gamma\left(-\frac{i\omega}{\lambda}-\alpha_j\right)\,.
\end{align}
The evenness condition $\Phi(\omega)=\Phi(-\omega)$ is automatically satisfied, but to ensure $\Phi(\omega)=\Phi(\omega)^*$ for real $\omega$, one needs that the set $\{\alpha_j\}_j$ is closed under complex conjugation. 

Now we analyse the small $\omega$ behaviour of this power spectrum. We again use that, as $\omega\to0$, $\Gamma(\omega-n) \sim (-1)^n/(n!\,\omega)$ for $n\in \mathbb Z_{\geq 0}$, to get
\begin{align}\label{eq:epsilon_approx2}
    \Phi (\omega) \sim \omega^{m-1-2k},\qquad \omega \to 0\,.
\end{align}
up to proportionality constant and where $k_{\text{pos.},0}$ is the number of zero and positive integer roots in the set $\{\alpha_j\}_j$. So, only for $m-1-2k \geq 0$ do we have a finite power spectrum at the origin. Note that, in contrast with~\eqref{eq:set_case1}, negative integer roots do not affect the regularity of the power spectrum at $\omega=0$.

Lastly, let us use the Weierstrass-type product formula to write the power spectrum as a product over its poles, 
\begin{align}\label{eq:Phi_case1_rational_earlier}
\Phi(\omega)= \pi^{-1} \left(\left( \omega/\lambda\right)^{m-1}\prod_{l=1}^\infty 	\left(1+\frac{ \omega^2}{\lambda^2 l^2}\right)\right)^{m-1} \prod_{j=1}^m |\Gamma(-\alpha_j)|^2 \prod_{n=0}^\infty \left( 1 + \frac{\omega^2}{\lambda^2 |n-\alpha_j |^2} \right)^{-1}\,.
\end{align}

\paragraph{Case 3.}
Finally, inspired by the example provided by the Schwarzian limit of SYK in eqs.~\eqref{eq:power_sp_schwarzian} and~\eqref{eq:G_Schwarzian}, let us consider the homogeneous solution
\begin{align}
	G(z)= 
i \sin\left(\frac{i\pi \omega}{\lambda}\right   )\prod_{j=1}^m	
\frac{1}
{\sin(\pi (\frac{i \omega}{\lambda}+\alpha_j))}\,.
\end{align}
The associated power spectrum is very similar to the previous case, and becomes
\begin{align}
  \Phi(\omega) 
  &= \pi^{-m}\prod_{j=1}^m	
\Gamma\left(\frac{i\omega}{\lambda}-\alpha_j\right)\Gamma\left(-\frac{i\omega}{\lambda}-\alpha_j\right)\,.
\end{align}
As for the previous case, the evenness condition $\Phi(\omega)=\Phi(-\omega)$ is automatically satisfied. To ensure $\Phi(\omega)=\Phi(\omega)^*$ for real $\omega$, the set of roots $\alpha_j$ must be closed under complex conjugation. 
When $m =1$ and $\alpha_1 = -\Delta$, we retrieve the SYK power spectrum~\eqref{eq:app_SYK_power_spectrum} up to a prefactor.
Because $\Delta>0$, there are no positive or zero roots in this set and thus, by the analysis in eq.~\eqref{eq:epsilon_approx2} with $m=1$, this case will lead to a finite power spectrum around $\omega=0$.

Finally, using the Weierstrass product expansion of the Gamma function, this power spectrum can be written as a product over its poles, 
\begin{align}\label{eq:Phi_case1_rational}
\Phi(\omega)= \pi^{-m} \prod_{j=1}^m |\Gamma(-\alpha_j)|^2 \prod_{n=0}^\infty \left( 1 + \frac{\omega^2}{\lambda^2 |n-\alpha_j |^2} \right)^{-1}\,.
\end{align}

\subsection{Relation to the Thermal Product Formula}\label{sec:rel_th_prod_form}

An interesting feature of the power spectra in  eq.~\eqref{eq:Phi_case1_rational} is that these are meromorphic functions of the form
\begin{align}\label{eq:power_spectrum_poles_str}
    \Phi(\omega) = \Phi(0)\prod_{n=1}^\infty \left(1 -\frac{\omega^2}{\omega_n^2} \right)^{-1} \left(1 - \frac{\omega^2}{\omega_n^{*2}} \right)^{-1}\,, \quad \Phi(0) > 0\,.
\end{align}
where  $\Phi(0)$ is a constant that only depends on the values of the poles and the set of poles as determined in the expression~\eqref{eq:Phi_case1_rational}. 
The poles are quasinormal modes (QNM) coming in quadruples $(\omega_n, -\omega_n,\omega_n^*,-\omega_n^*)$, where $\mathrm{Im}\, \omega_n < 0$. For the example in eq.~\eqref{eq:Phi_case1_rational}, these frequencies $\omega_n$ are identified with the roots of $p(z)$, namely they are~$i\lambda(n-\alpha_n)$. 

The general expression in eq.~\eqref{eq:power_spectrum_poles_str} was conjectured by positing a minimal set of constraints on the power spectrum, motivated by thermal conditions and assuming conformal symmetry and holography. In particular, besides the general properties of unitarity and KMS, \cite{Dodelson:2023vrw} demanded that the power spectrum 1) only display singularities at isolated \textit{simple} poles in the complex plane, 2) not possess any zeroes (i.e $1/\Phi(\omega)$ is an entire function), and 3) a constraint on the asymptotic behaviour requiring that along any ray in the complex plane that asymptotically avoids poles, the power spectrum satisfies
\begin{align}\label{eq:rayconstraint}
\frac{1}{\left|\Phi(\omega)\right|} 
    \sim e^{\lambda(\theta) r}\,,
    \quad 
    \omega = r e^{i \theta} \to \infty\,,
    \quad
    \lambda(\theta) \ge 0\,,
\end{align}
up to a polynomial prefactor. These conditions are motivated by a combination of holographic and thermal arguments.  

Earlier, in eq.~\eqref{eq:Phi_case1_rational}, we found solutions to the shift symmetry constraints that bear a strong resemblance to the thermal product formula~\eqref{eq:power_spectrum_poles_str} of \cite{Dodelson:2023vrw}. 
There are, however, several important differences:
\begin{itemize}
    \item In the construction of \cite{Dodelson:2023vrw}, the two-sided thermal correlator is obtained by solving a wave equation in a black hole background, leading to a power spectrum where all singularities are isolated and \textit{simple} poles. These poles correspond to QNMs of the black hole geometry and encode the real-time decay of thermal perturbations. The assumption of simplicity is therefore not an auxiliary condition but intrinsic to the physical origin of the correlator.

In contrast, the power spectrum we derive, based on thermal principles and maximal chaos in the out-of-time-order correlator, need not satisfy this restriction. The resulting function in eq.~\eqref{eq:Phi_case1_rational} is meromorphic but may exhibit poles of higher degree. These singularities do not necessarily correspond to QNMs in the gravitational sense and reflect a different analytic structure that goes beyond the assumptions of the black hole setup.
    \item In \cite{Dodelson:2023vrw}, the power spectrum did not possess any zeros in the complex $\omega$-plane, implying that $1/\Phi(\omega)$ is an entire function. This non-vanishing property arises from solving the bulk wave equation in a smooth black hole background, where the ingoing boundary condition at the horizon and the analytic structure of the scattering potential ensure a zero-free correlator \cite{Dodelson:2023vrw}. In contrast, our (holography-inspired) construction derives the power spectrum from general thermal principles together with maximal chaos in the out-of-time-order correlator. As seen, for example, in eq.~\eqref{eq:nonsimpl_case1}, power spectra derived from the shift symmetry constraints that satisfy a minimal set of physical conditions can vanish at $\omega=0$. 
    \item Finally, turning to the requirement in eq.~\eqref{eq:rayconstraint}, since at large frequencies the spectra in eq.~\eqref{eq:Phi_case1_rational} behaves as
\begin{align}\label{eq:decay_sols}
    \Phi(\omega) \sim e^{-\pi \omega/\lambda} \left( \frac{\omega}{\lambda} \right)^{-2\sum_j \alpha_j - m}\,,
\end{align}
this condition is automatically fulfilled. 
\end{itemize}

One has to keep in mind, though, that the power spectrum $\Phi(\omega)$ in eq.~\eqref{eq:power_spectrum_poles_str} pertains only to the bare field $\hat V$. 
We have the working assumption that the two-point function for $\hat V$ provides a reliable approximation to that of the total field $V$, i.e. the corrections introduced by $\epsilon$ are small and can be truncated.

\section{Discussion}\label{sec:discussion}

In this work, we contrasted prominent diagnostics of quantum chaos, the OTOC and Krylov space-based probes like Krylov complexity and the UOGH, in an EFT describing maximally chaotic quantum systems.
In the EFT construction, the exponential growth of OTOCs with the maximal Lyapunov exponent arises due to 
a shift symmetry \cite{Blake:2017ris, Blake:2021wqj}, which also imposes a constraint on the two-point functions of simple operators. A natural question is then whether this same structure is strong enough to enforce maximal Krylov growth. Motivated by the conjectured relation~\eqref{eq:conjecture}, which suggests that $\lambda_K$ should be bounded by the chaos scale and saturates $2\pi T$ in maximally chaotic systems, we explored this question within the EFT. To this end, we constructed families of thermal two-point functions consistent with the shift symmetry constraints as well as a minimal set of physical constraints appropriate for thermal systems, including KMS symmetry, unitarity, and reality from Hermiticity of operators. We computed the corresponding Lanczos coefficients, and, within the set of two-point functions for which we could analytically calculate the Lanczos coefficients, we did indeed find some with $\lambda_K = 2\pi T$, but also some with $\lambda_K = \pi T$  

Next, we analysed the general structure of solutions to the shift symmetry equation after imposing the minimal set of physical constraints.
In the time domain, the correlator solutions are Meijer G-functions.
In the frequency domain, a subset of our solutions resemble the ``thermal product formula'' of \cite{Dodelson:2023vrw}, sharing the same overall structure but differing in important analytic features: the holographic power spectra have only simple poles - no poles of order $n >1$ - and no zeroes, whereas power spectra in the EFT allow for more general singularity structures, including both higher-order poles and zeroes.

\subsubsection*{Krylov exponents and EFT of maximal chaos}
In this work, we constructed explicit examples of thermal two-point functions consistent with physical and shift symmetry constraints in the EFT, and used them to study the behaviour of Krylov exponents. In~\autoref{sec:Krylov_vs_EFT}, we found that such correlators can give rise to both maximal and submaximal Krylov growth, demonstrating that saturation of the MSS bound, $\lambda_L \leq 2\pi T$, does not imply maximal Krylov exponent.
So, the shift symmetry in the EFT is sufficient to ensure maximal Lyapunov exponent, but not maximal Krylov exponent.

While we were able to solve the finite $m$ shift symmetry constraint equation in generality, we were only able to analytically calculate the Lanczos coefficients and Krylov exponents for a subset. 
Interestingly, for all correlators in that subset, the Lanczos coefficients $b_n$ are asymptotically linear and thus satisfy the UOGH.
However, the full set of solutions consistent with shift symmetry is vast, and we cannot exclude the possibility of more exotic behaviours. Further work on this, both analytical and numerical, is warranted. A natural next step would be to determine what analytic structure in the two-point function gives rise to non-smooth $b_n$, including staggering, and to determine whether such solutions are physically realised within the constraints of the EFT. Early explorations in this direction can be found in \cite{Camargo:2022rnt,Dymarsky:2021bjq,Avdoshkin:2022xuw}.

We also highlighted that the Krylov exponent, unlike the Lyapunov exponent, is sensitive to UV features of the theory. 
While a given instance of the EFT is only expected to be an accurate approximation of the microscopic theory for timescales between relaxation and scrambling, when UV modes have already decayed, the Lanczos coefficients are constructed from spectral moments that receive contributions from high-frequency tails. 
This makes the Krylov exponent $\lambda_K$ in the microscopic theory potentially sensitive to UV physics beyond the reach of the EFT.

Our analysis is at leading order in $1/N$, and there is currently no formulation of the EFT that captures subleading corrections due to difficulties in systematically implementing KMS symmetry beyond leading order~\cite{Crossley:2015evo, Glorioso:2017fpd}. Should such a formulation become tractable, it would enable a more refined analysis of thermal correlators and potentially sharpen the EFT predictions for  Krylov space-based quantities and their comparison to the UOGH.

Our results also open the door to several directions for generalisations:
\begin{itemize}
    \item Other EFT frameworks, such as the effective descriptions developed in \cite{Haehl:2018izb}, where conformal symmetry plays a central role. Understanding whether our findings map onto theirs, and what distinguishes truly universal chaotic behaviour from model-specific features,  could clarify the extent to which the SYK model, with its Schwarzian and large $q$ limits, is a typical or exceptional chaotic theory.
\item Our focus is on one-dimensional systems, and extending these ideas to higher dimensions is an important challenge. The construction of a systematic EFT for operator dynamics in higher-dimensional chaotic systems is still lacking, though recent progress \cite{Knysh:2024asf} has opened new possibilities.
\item We have restricted our attention to maximally chaotic systems, but a recent proposal \cite{Gao:2023wun} has extended the EFT approach to systems with submaximal chaos. Developing tools to probe Krylov exponents in this class of thermalising quantum systems is another interesting next step.
\end{itemize}

\subsubsection*{Solutions to the shift symmetry constraints and the thermal product formula }
In the second part of this work, we investigated the general structure of thermal correlator solutions of the shift symmetry constraints.
We solved the arbitrary but finite order shift symmetry constraint equation in generality, and imposed physical constraints, to construct a large subset of all thermal two-point functions allowed for within the EFT of maximal chaos. 

Within the general set of solutions, given in eq.~\eqref{eq:solsPhi}, for a particular choice of periodic factor $G$ inspired by the conformal limit of SYK, we found power spectra that closely resemble the thermal product formula introduced in \cite{Dodelson:2023vrw}, and further developed in \cite{Bhattacharya:2025vyi}. In these examples, this structure arises from the structure of the general solution, which follows from the shift symmetry constraints and physical requirements, and our choice of $G$, without assuming a specific holographic dual.  

We also identified several other families of solutions within the general construction that violate the assumptions of having no zeroes and simple poles underlying the thermal product formula of \cite{Dodelson:2023vrw}. This provides an interesting contrast: our approach is based solely on imposing maximal chaos and the minimal conditions required for a consistent thermal system, rather than the stricter assumptions used in the thermal product derivation. While no explicit holographic dual is assumed, the effective field theory framework is nonetheless strongly guided and motivated by holographic intuition.

Our approach parallels several recent developments in thermal CFTs \cite{Iliesiu:2018fao,Marchetto:2023xap, Barrat:2025wbi, Barrat:2025nvu} and holographic correlators \cite{Buric:2025anb}, where thermal two-point functions are derived from minimal analytic, OPE, and symmetry input, supplemented by KMS invariance. Remarkably, such basic ingredients suffice to reconstruct non-trivial thermal observables. Exploring how far methods based on extracting thermal systems from consistency conditions can be pushed, and clarifying their limitations, is key to understanding the interplay between universal features and model‑dependent aspects of thermal physics.

\acknowledgments
We would like to thank Mike Blake, Hugo Camargo, Rathindra Nath Das, Marine De Clerck, Anatoly Dymarsky, Oleg Evnin, Oleg Lychovskiy, Enrico Marchetto, Alessio Miscioscia, Natalia Pinzani-Fokeeva and Adri\'an S\'anchez-Garrido for useful comments and discussions. We are also grateful to Shira Chapman and Ben Craps for input at the onset of this project and comments on the draft. The work of SD is supported by an Azrieli fellowship funded by the Azrieli foundation, and is partially supported by the Israel Science Foundation (grant No. 1417/21), by Carole and Marcus Weinstein through the BGU Presidential Faculty Recruitment Fund, the ISF Center of Excellence for theoretical high energy physics and by the German-Israel-Project (DIP) on Holography and the Swampland, and by the ERC Starting Grant dSHologQI (project number 101117338). The work of SD is also supported in part by Deutsche Forschungsgemeinschaft under Germany's Excellence Strategy EXC 2121 Quantum Universe 390833306. SD wishes to thank DESY and Universit\"at Hamburg for the kind hospitality. The work of MK and AR is supported by FWO-Vlaanderen project G012222N and by the VUB Research Council through the Strategic Research Program High-Energy Physics. The work of AR is also supported by the FWO-Vlaanderen through a Senior Postdoctoral Fellowship 1223125N.

\appendix

\section{Recursion method for calculating Lanczos coefficients} \label{app:toda}
A method to calculate Lanczos coefficients through a simple recursion relation is given in~\cite{Dymarsky:2019elm}.
Given a thermal correlator $C(\tau)$, with $\tau$ Euclidean time, and regulated ($C(\tau):=\Tr{(\rho_{\beta}^{1/2} e^{-H \tau} V(0) e^{H \tau} \rho_\beta^{1/2} V(0)}$), one can find the Lanczos coefficients by solving the recursion relation
\bne\label{eq:dym} 
\frac{d^2}{d\tau^2} \log b_{n}(\tau)^2 = b_{n+1}(\tau)^2 - 2 b_n(\tau)^2 + b_{n-1}(\tau)^2\,, \quad  b_{-1}(\tau) = 0\,, \; b_0 (\tau)^2 = \frac{d^2}{d\tau^2} \log C(\tau)\,, 
\ene
and the Lanczos coefficients are $b_n = b_n (0)$.  This equation is related to the Toda equation, which is integrable, so in principle, there is always a solution. One simple consequence of this equation is that for solutions of the type $b_{n} (\tau)^2 = b(\tau)^2 p(n)$, the polynomial in $p(n)$ must be quadratic in $n$. The recursion relation is simple to solve in this case.

\section{Physical properties of thermal autocorrelators}\label{app:phys_props}
Here we summarise key physical properties of thermal autocorrelators. For more on these properties, as well as general background on thermal field theory, see the textbooks~\cite{Bellac:2011kqa, Das:1997gg, Laine:2016hma, Kapusta:2006pm}.

We will only discuss bosonic operators here, but analogous results exist for fermionic operators with appropriate modifications. See~\autoref{eq:KC_UOGH} for definitions of $C$ and $\Phi$.
\begin{itemize}
    \item {\bf KMS condition.}
    The Kubo-Martin-Schwinger (KMS) condition characterises thermal equilibrium and implies a specific periodicity structure in imaginary time
    \begin{align}
    	g_{\hat V}(t + i\beta) = g_{\hat V}(-t) \quad \Leftrightarrow \quad C(t) = C(-t)\,,
    \end{align}
   for the KMS-symmetric autocorrelator $ C(t) = g_{\hat V}(t + i\beta/2) $. The KMS condition applied to the power spectrum reads $\Phi(\omega)=\Phi(-\omega)$.

    \item {\bf Reflection positivity and unitarity.}
    Reflection positivity is a Euclidean signature condition that is important for unitarity in Lorentzian signature via the Osterwalder-Schrader reconstruction theorem. For our two-point functions, it requires
    \begin{align}
    g_{\hat V}(i\tau) \geq 0 \quad \text{for all } \tau \in [0, \beta]\,,
    \end{align}
    for bosonic operators. This ensures that the Euclidean correlator defines a positive norm on the Hilbert space.

    In the frequency domain, this implies that the power spectrum is real and non-negative
    \begin{align}
    \Phi(\omega) \geq 0, \qquad \omega \in \mathbb{R}\,.
   \end{align}

    \item {\bf Thermal decay.}
    At large time separation, thermal correlators decay
    \begin{align}
    g_{\hat V}(t) \sim e^{-|t|/t_c}, \qquad |t| \to \infty\,,
    \end{align}
    where $ t_c $ is the thermal correlation time. For thermal CFTs, this is controlled by the operator dimension, as the two-point function reads
     \begin{align}
    g_{\hat V}(t) = \frac{1}{\sinh^{2\Delta} (\pi t/\beta)} \sim e^{-2\pi \Delta t/\beta}, \qquad t \to \infty,
    \end{align}
    hence $t_c = \beta / (2\pi \Delta)$.

    The decay rate determines the analyticity of the Fourier transform $\Phi(\omega)$. If  $g_{\hat V}(t) \sim e^{-|t|/t_c}$, then $\Phi(\omega)$ is analytic in the strip
     \begin{align}
    |\text{Im}(\omega)| < \frac{1}{t_c}\,.
    \end{align}
     \item {\bf Paley-Wiener decay bound for thermal spectra.}
For a quantum system in thermal equilibrium at inverse temperature $\beta$, the  Wightman function $C(t)$ must be analytic in the complex-time strip $|\mathrm{Im}(t)| < \beta/2$, as required by the KMS condition. This analyticity imposes a constraint on the asymptotic behaviour of the power spectrum $\Phi(\omega)$, that is
\begin{align}
    \Phi(\omega) \lesssim e^{-\delta |\omega|}\,,
\end{align}
for some $\delta \ge \beta/2$.
This is a direct consequence of the Paley-Wiener theorem: exponential decay of $\Phi(\omega)$ is necessary to ensure strip analyticity of $C(t)$.
If $\Phi(\omega)$ decays slower than exponentially (e.g. $\Phi(\omega) \sim \omega^{-p}$), then the inverse Fourier transform $C(t)$ fails to be analytic in any non-trivial strip around the real axis. 
\end{itemize}

\section{Time domain solutions using  Meijer's G-functions}\label{App:Meijers_hypergeom}

This appendix compiles definitions and properties of the generalised hyperbolic and Meijer's G-function. In addition, we include detailed relations and derivations, completing the analysis in the main text. Next, we use these to identify the shift symmetry constraints with special cases of the  Meijer G-function differential equations.
\subsection{Defining the Meijer G-function}
The Meijer G-function (see chapter 9.3  of \cite{gradshteyn2014table}, sec. 5.3 in volume I of \cite{erdelyi1953htf1} or chapter 16 of \cite{NIST:DLMF}) can be defined through its Mellin-Barnes representation 
\begin{align}\label{eq:MeijerGfct}
   G^{m,\,n}_{p,\,q} \left( z \,\Bigg|\, \begin{array}{c} a_1, \dots, a_p \\ b_1, \dots, b_q \end{array} \right)=
\frac{1}{2\pi i}\int_{C}z^{s}
\,\frac{\prod_{j=1}^{m}\,\Gamma(b_j - s)\;}
{\prod_{j=1}^{n}\,\Gamma(a_j - s)}
\frac{\prod_{j=m+1}^{q}\,\Gamma(1 - b_j + s)}
{\prod_{j=n+1}^{p}\,\Gamma(1 - a_j + s)}\,\mathrm ds\,,
\end{align}
where $m$ and $n$ are integers such that
$0 \leq m \leq q$ and~$\quad 0 \leq n \leq p$ as well as none of $a_k - b_j$ is a positive integer when
$1 \leq k \leq n$ and $1 \leq j \leq m$. 

There are three possible choices for the integration contour $C$: 
\begin{itemize}
    \item[(i)] $C$ goes from $-i\infty$ to $i\infty$. The integral converges if $p + q < 2(m + n)$ and~$|\arg z|<\left( m + n - \frac{1}{2}(p + q) \right)\pi$. We will refer to this type of contour as the Bromwich-type.\label{contour1}
    \item[(ii)] $C$ is a loop that starts at infinity on a line parallel to the positive real axis, encircles counter-clockwise the poles of the $\Gamma(b_\ell - s)$ once and returns to infinity on another line parallel to the positive real axis. The integral converges for all $z \neq 0$ if $p < q$, and for $0 < |z| < 1$ if $p = q \geq 1$. \label{contour2}
  \item[(iii)] $C$ is a loop that starts at infinity on a line parallel to the negative real axis, encircles the poles of the $\Gamma(1 - a_\ell + s)$ once in the positive sense and returns to infinity on another line parallel to the negative real axis. The integral converges for all $z$ if $p > q$, and for $|z| > 1$ if $p = q \geq 1$. \label{contour3}
\end{itemize}

Equivalently, one can characterise Meijer's G-functions as solutions to the differential equation
\begin{align}\label{eq:diff_Meijers}
    \left[
(-1)^p x \prod_{j=1}^p \left( D - a_j + 1 \right)
	-	\prod_{j=1}^q \left( D - b_j \right)
\right] f(x) = 0\,,
\end{align}
where $D = x \frac{\mathrm d}{\mathrm d x}$. The Meijer function is related to hypergeometric functions through Slater's theorem; eq. (16.17.2) in \cite{NIST:DLMF};  
\begin{align}\label{eq:slater}
G^{m,n}_{p,q} \left(
\begin{array}{c}
a_1, \dots, a_p \\
b_1, \dots, b_q
\end{array}
\biggm| z \right)
&= \sum_{k=1}^{m} A^{m,n}_{p,q,k}(z)\;
{}_{p}F_{q-1} \left(
\begin{array}{c}
1 + b_k - a_1, \dots, 1 + b_k - a_p \\
1 + b_k - b_1, \dots, \star, \dots, 1 + b_k - b_q
\end{array}
\Bigm| (-1)^{p - m - n} z \right)\,,
\end{align}
where the $\star$ indicates omission of the argument involving $b_k$, i.e. $1 + b_k - b_k=1$, and 
\begin{align}\label{eq:A_slater}
A^{m,n}_{p,q,k}(z) :=
\frac{z^{b_k}
\prod_{\substack{j=1\\ j \ne k}}^{m} \Gamma(b_j - b_k)
\prod_{j=1}^{n} \Gamma(1 + b_k - a_j)}
{\prod_{j = m+1}^{q} \Gamma(1 + b_k - b_j)
\prod_{j = n+1}^{p} \Gamma(a_j - b_k)}\,.
\end{align}

A special subclass of functions that the Meijer G-functions describe are generalised hypergeometric functions. A useful form of the defining relation for the hypergeometric function is 
\begin{align}
    F(a, b; c; z) = \sum_{s=0}^{\infty} \frac{(a)_s (b)_s}{(c)_s s!} z^s
= \frac{\Gamma(c)}{\Gamma(a) \Gamma(b)} \sum_{s=0}^{\infty}
\frac{\Gamma(a + s)\, \Gamma(b + s)}{\Gamma(c + s)\, s!} z^s\,,
\end{align}
where $(\alpha)_n$ are Pochhammer symbols.

A Fourier transform used in \autoref{sec:anti_per_KMS} is 
 \begin{align}\label{eq:FT_sinh/sinh}
     \int_{-\infty}^{\infty} \frac{\sinh(a t)}{\sinh(b t)} e^{-i \omega t} dt = \frac{\pi}{b} \frac{\sin\left( \frac{\pi a}{b} \right)}{\cosh\left( \frac{\pi \omega}{b} \right) + \cos\left( \frac{\pi a}{b} \right)},
 \quad \text{for }\; 0 < a < b\,,
 \end{align} 
 see \cite{gradshteyn2014table} Table 17.23, instance $20$, and
\begin{align}\label{eq:gamma_coshcos}
    \Gamma(1 + x + i y)\, \Gamma(1 - x + i y)\, \Gamma(1 + x - i y)\, \Gamma(1 - x - i y)
= \frac{2 \pi^2 (x^2 + y^2)}{\cosh(2\pi y) - \cos(2\pi x)}\,,
\end{align}
for  $x, y \in \mathbb{R}$, see eq. (8.332.4) in \cite{gradshteyn2014table}.

\subsection{Meijer's G-function realisation of the solutions}\label{app:alternative_via_diff}
An alternative route to solving the shift symmetry constraints is to directly identify them with the defining differential equations of special functions. In this section, we outline this method. For concreteness, we will restrict the analysis to the $(+\lambda)$-shift symmetry constraint.

As a warm-up, let us consider the case where the infinite sum in eq.~\eqref{eq:shifteq} truncates at order two. The shift symmetry constraint then takes on the form
	\begin{align}
	(1 + e^{t \lambda}) f_0(\lambda) g_{\hat V}(t) 
+ (-1 + e^{t \lambda}) f_1(\lambda) \partial_t g_{\hat V}(t) 
+ (1 + e^{t \lambda}) f_2(\lambda) \partial_t^2 g_{\hat V}(t) = 0\,.
\end{align}
Solutions to this equation are built out of the hypergeometric series ${}_2F_1(a,b,c,z)$ with $a,b$ and $c$ being parameters defined in terms of $f_0,f_1$ and $f_2$ and $z=-e^{\lambda t}$. An example can be found in eq.~\eqref{eq:m2_autocorr}. In particular, the solutions are $g_{\hat V}(t)= e^{-Bt/2}w(z)$  where $B=(-f_1\pm\sqrt{f_1^2-4f_0f_2})/f_2$. The functions $w(z)$ are solutions to the hypergeometric differential equation
	\begin{align}\label{eq:hypergeom}
z(1-z)\frac{\mathrm d^2w}{dz^2} + \left(c-(a+b+1)z\right)\frac{\mathrm dw}{dz} - ab\,w = 0\,.
	\end{align}
More generally, the higher-order shift symmetry constraint is solved in terms of generalised hypergeometric functions via Meijer's G-functions. Slightly rewriting the differential equations in eq.~\eqref{eq:diff_Meijers}, consider 
\begin{align}
   \Big[ (-1)^{p-m-n}\, z\,
\bigl(D - a_1 + 1 \bigr)\cdots \bigl(D - a_p + 1 \bigr)-
\bigl(D - b_1 \bigr)\cdots \bigl(D - b_q \bigr)
\Big]\, f(z) = 0\,,
\end{align}
where $D=z\frac{\mathrm d}{\mathrm d z}$ and $f(z)$ is the Meijer's G-functions. 
As will become clear soon, let us consider the particular class of Meijer's functions of the form
\begin{align}
    G^{m,1}_{m,m} \left( z \,\Bigg|\, \begin{array}{c} 0,\ 1 + \alpha_1,\dots, 1 + \alpha_{m-1} \\ 1,\ -\alpha_1,\dots, -\alpha_{m-1} \end{array} \right)\,,
\end{align}
which solves the specialised differential equation
\begin{align}\label{eq:M_diffeq_class1}
z (D + 1) \prod_{j=1}^{m-1} (D - \alpha_j)\, f(z)
= (D - 1) \prod_{j=1}^{m-1}(D + \alpha_j)\, f(z)\,.
\end{align} 
Defining $\gamma_l=(-1,\alpha_1,\dots)_l$ leads to
\begin{align}\label{eq:M_diffeq_class1_v2}
    z \prod_{l=0}^{m-1} (D - \gamma_l) f(z) - \prod_{l=0}^{m-1} (D + \gamma_l) f(z)= 0\,.
\end{align}
Combining the terms and expanding the sum leads to
\begin{align}\label{eq:sum_Gmmm1}
    \sum_{l=0}^{m}c_l(z+(-1)^l)D^lf(z)=0\,.
\end{align}
where $\prod_{l=0}^{m-1} (D \pm \gamma_l)=\sum_{l=0}^{m} (\pm 1)^lc_l\,D^l$ and the coefficients $c_l$ are determined in terms of elementary symmetric functions of the parameters in the Meijer G-function: $c_l = e_{m+1-l}\bigl(-1,\alpha_1,\dots,\alpha_m\bigr)$. Turning back to the shift symmetry constraint~\eqref{eq:shift}, and performing the transformation $z=-e^{\lambda t}$ and identifying $g_{\hat V}(t)$ with $f(z)$, leads to
\begin{align}\label{eq:shift_z_eq}
\sum_{n=0}^m f_n(\lambda)\, (z + (-1)^n)\, (\lambda D)^n  g_{\hat V}(t(z)) = 0\,,
\end{align}
upon the identification $f_n(\lambda)=\,c_n\lambda^n$ this is equivalent to eq.~\eqref{eq:sum_Gmmm1}. Similarly, one can derive the differential equation defining $G^{2m,0}_{0,2m}$-Mejier's G-function.


\section{General solution in the time domain for (anti)-periodic power spectra}
\label{app:nonsimpleroots}
In this appendix, we expand on the cases left out of the analysis in  \autoref{sec:anti_per_KMS}, considering the general set of solutions to the specialised shift equation in eq.~\eqref{eq:shifp_id}, without imposing any further conditions.

If the roots $\alpha_i$ are all simple, the general solution is~\eqref{eq:genso2_C}.
When the polynomial has non-simple poles $p(z) = \lambda^m f_m(\lambda)\prod_{i=1}^r (z - \alpha_i)^{m_i}$, where $r$ labels the distinct roots $ \alpha_1, \dots, \alpha_r$, and each root $\alpha_i$ has multiplicity $m_i$ such that $m = \sum_{i=1}^r m_i$, the solution to the eq.~\eqref{eq:shifp_id} becomes
\begin{align} \label{eq:spec2}
g_{\hat V}(t) = \frac{ \sum_{i=1}^r \sum_{k=0}^{m_i - 1} c_{i,k} t^k e^{-\alpha_i \lambda t} }{ 1 + (-1)^m e^{\lambda t} }\,.
\end{align}
This is valid for arbitrary complex $\alpha_i$, though the thermal decay property requires Re$(\alpha_i) > -1$. If one of the roots is zero, then we get a term in the numerator of~\eqref{eq:spec2} that is purely a polynomial.

The solutions in eq.~\eqref{eq:spec2} differ from those in~\eqref{eq:genso2_C} by additional powers of $t$ in the numerator, so, using that the power spectrum of $t^n C(t)$ is $(-i)^n \frac{d^n}{d\omega^n} \Phi(\omega)$, the structural form of the power spectra of the general correlator~\eqref{eq:spec2} will include derivatives of the power spectra in~\eqref{eq:Phi_spec_case_noKMS}.

\bibliography{references}
\bibliographystyle{JHEP}
\end{document}